\documentclass[]{tCPH2e}

\usepackage{epstopdf}
\usepackage{subfigure}
\usepackage{eurosym}
\usepackage[]{algorithm2e}

\usepackage{booktabs}
\usepackage[dvipsnames]{xcolor}
\usepackage{tikz}

\usepackage{array}
\usepackage{calc}

\usepackage{multirow}

\usepackage{amsthm}

\theoremstyle{plain}
\newtheorem{theorem}{Theorem}[section]

\theoremstyle{definition}
\newtheorem{definition}[theorem]{Definition}
\newtheorem{example}[theorem]{Example}

\theoremstyle{remark}

\usepackage{geometry}                		
\usepackage{graphicx}				
\usepackage{amssymb}	
\usepackage{color} 		
\usepackage[numbers,sort&compress]{natbib}
\bibpunct[, ]{[}{]}{,}{n}{,}{,}
\makeatletter
\def\NAT@def@citea{\def\@citea{\NAT@separator}}
\makeatother

\newcommand{\dummyfigure}[2]{\includegraphics[height=#2 cm]{#1}}

\newtheorem{prob}{Problem}

\begin{document}
\articletype{}

\title{A cross-disciplinary introduction to quantum annealing-based algorithms}

\author{
\name{Salvador E. Venegas-Andraca\textsuperscript{a}\thanks{CONTACT S.E. Venegas-Andraca. Email: salvador.venegas-andraca@keble.oxon.org, salvador@venegas-andraca.org}, William Cruz-Santos\textsuperscript{b}, Catherine McGeoch\textsuperscript{c}, and Marco Lanzagorta\textsuperscript{d}}
\affil{\textsuperscript{a}Tecnologico de Monterrey, Escuela de Ingenieria y Ciencias. Ave. Eugenio Garza Sada 2501, Monterrey, NL 64849, Mexico; \\ \textsuperscript{b}CU-UAEM Valle de Chalco, Estado de M\'exico, M\'exico;\\ \textsuperscript{c}D-Wave Systems, 3033 Beta Avenue, Burnaby, British Columbia, V5G4M9, Canada.\\ \textsuperscript{d}US Naval Research Laboratory, 4555 Overlook Ave. SW, Washington DC, 20375, USA.}
}

\maketitle

\begin{abstract}
A central goal in quantum computing is the development of quantum hardware and quantum algorithms in order to analyse challenging scientific and engineering problems.  Research in quantum computation involves contributions from both physics and computer science, hence this article presents a concise introduction to basic concepts from both fields that are used in annealing-based quantum computation, an alternative to the more familiar quantum gate model. 

We introduce some concepts from computer science required to define difficult computational problems and to realise the potential relevance of quantum algorithms to find novel solutions to those problems. We introduce the structure of quantum annealing-based algorithms as well as two examples of this kind of algorithms for solving instances of the max-SAT and Minimum Multicut problems. An overview of the quantum annealing systems manufactured by D-Wave Systems is also presented.
\end{abstract}


\section{Introduction}
\label{introduction}

Cross-pollination between physics and computer science has long been mutually beneficial. For example, the development of semiconductor-based computer technology would be unthinkable without solid-state physics \cite{yu2010,itrs2015,wong2002,wong2010,avci2014}; similarly, progress on advanced algorithms has benefited from mathematical models of physical phenomena \cite{kirkpatrick83,baez12}. As for the influence of computer science, a novel approach in scientific research is to  model natural phenomena using the tools of theoretical computer science \cite{compeau12,cubitt15}. 

Quantum computation is a scientific and engineering field focused on developing information processing devices and algorithms based on quantum mechanics. In addition to further advancing the theoretical and experimental foundations of this discipline  and the emergence of fields like quantum machine learning \cite{wittek14,schulda15,biamonte17} and quantum image processing  \cite{yan15,aburaed17}, quantum computing and the broader field of quantum technologies (embracing computing, communications, cryptography and sensing) are  becoming an attractive emerging branch of high-tech business    
\cite{idquantique,dwave,ibm,microsoft,googlequantum,1qbit,rigetti,qmanifesto,ukquantumhubs,qis_executive_usa,mittechreviewquantum2017,winiarczyk13,ukquantumtech14,usclockheedmartin,hpquantumlabs}.


Due to the asynchronous progress of theoretical and experimental quantum computing, most quantum algorithms have been designed on paper and, sometimes, tested on simulation software (recent examples include \cite{playground,liquid,qmdirac}). The advancement of quantum computing as a scientific and engineering discipline requires the physical realisation of quantum hardware on which quantum algorithms can be tested.

\newpage{}

Quantum annealing is a physical platform of quantum computation focused on solving combinatorial optimisation problems. Quantum annealing is a restricted form of adiabatic quantum computation but the problems that can be explored using this paradigm are vast and relevant to many fields of science and technology. D-Wave systems manufactures commercially available quantum annealing-based hardware on which it is possible to run algorithms. Thanks to the access granted by USRA-NASA to the D-Wave's quantum annealer installed at NASA Ames Research Centre, our work on quantum annealing-based algorithms and the examples presented in this paper have been designed, tested and run using both D-Wave's quantum annealer and D-Wave's advanced simulation software. Future quantum annealers probably will use similar hardware and programming principles as D-Wave's, so we expect that our discussions and results will have a broad impact in the field of annealing-based quantum computation.


In this paper, we present a cross-disciplinary introduction to quantum annealing-based computer technology and algorithms. Our purpose is to present a succinct yet complete introduction of this field with emphasis on the continuous exchange of paradigms and tools between physics and computer science.  Section \ref{computerscience} presents key computer science concepts which are needed to understand the nature and relevance of quantum annealing-based algorithms, section \ref{dwavecathy} introduces fundamental components and properties of D-Wave's quantum annealing hardware, section \ref{william} presents detailed examples of quantum annealing-based algorithms designed to run on D-Wave's hardware, and section \ref{conclusions} presents some conclusions.

\section{Introduction to computer science and annealing-based algorithms}
\label{computerscience}

This section succinctly introduces  automata, computational complexity, simulated annealing and quantum annealing. Automata and computational complexity are essential tools to understand the mathematical properties of computers and algorithms, while classical and quantum annealing are computational methods devised to probabilistically solve  problems efficiently that would otherwise require a colossal amount of computational resources.


\subsection{Automata and (non)-deterministic computation}

\subsubsection{Automata}

\begin{figure}[hbt]
\begin{center}
\scalebox{0.4}{\includegraphics{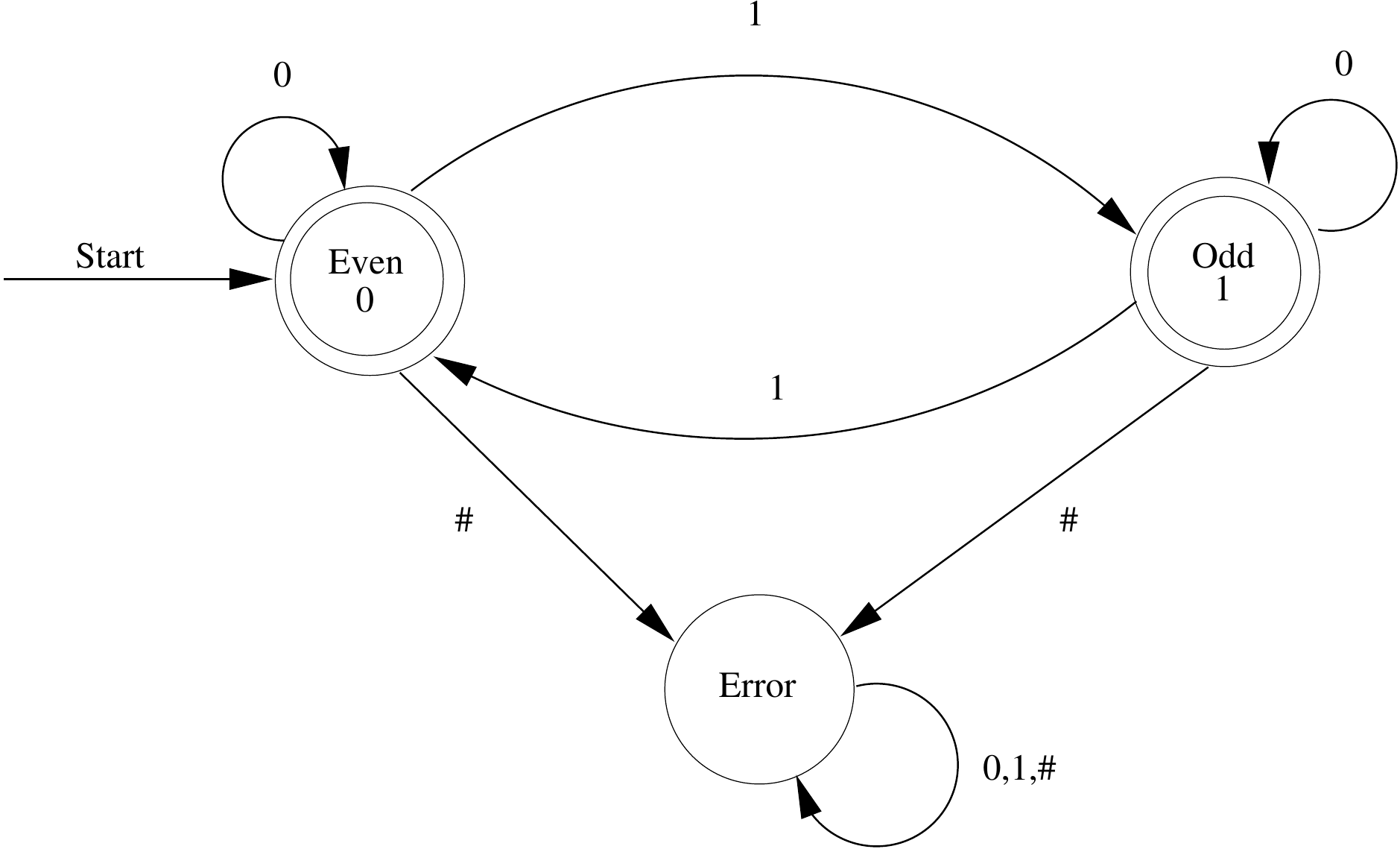}}
\end{center}
\caption{{\small A finite automaton $\cal{R}$ for parity computation.
Double-circled states are accept states and single-circled state
is a reject state. Input strings for $\cal{R}$ are any combination of characters taken from 
$\Sigma = \{0,1,\# \}$. 
Since an empty set of characters has no $1$s, we set the initial parity of a string as even (this is why the state \lq Even' has an arrow labeled as \lq Start').
}
}\label{automaton_parity}
\end{figure}

\newpage{}

Automata are mathematical models of physical devices used to compute. The definition of any automaton requires three components:  states, language and transition rules.  States are abstractions of the different stages an automaton would go through while computing (e.g., {\it on} and {\it off} states of a switch).
A language is a set of strings made by concatenating characters from a given alphabet. For example, one could define a language from alphabet $\{$a,b$\}$,  consisting of all strings that begin with \lq a'  and end with \lq b'.   This language would include  \lq ab', \lq aabb', and  \lq ababab',  but not \lq ba' or \lq bbb'.   Finally, transition rules describe the internal dynamics of an automaton (i.e., how and when an automaton changes its state).


Computation in an automaton is performed by following sequences of operations (i.e. moving through states) determined by the transition rules of the automaton and the input string fed to the automaton. The final state of a computation performed by an automaton is an {\it accept} ({\it reject}) state if the input string belongs (does not belong) to the language $L$ upon which the automaton has been defined.

For instance, we introduce automaton $\cal{R}$ (Fig.~(\ref{automaton_parity})). $\cal{R}$ has three states (denoted by circles), several transition rules (denoted by arrows) and language $L = \{\omega|\omega \text{ is a string composed of 0s and 1s}\}$. $\cal{R}$, located at one of the ends of a digital communication system, is used to compute the parity of  a string. Input for $\cal{R}$ is a string taken from $\{ 0,1,\# \}$, where  \lq $\#$'  is used to state that a transmission error has occurred. The parity of a string  is odd (even) if its number of $1$s is odd (even). Only strings with no errors are suitable for parity computation, hence only strings of $0$s and $1$s will be accepted and any string with at least one $\#$ shall be rejected. 

Let us show the behavior of $\cal{R}$ with two examples. First, let $\cal{R}$
receive string $100110001$ as input (read from left to right).
The first input character is \lq $1$' thus $\cal{R}$ goes from state
\lq Even' to state \lq Odd'. We then read \lq 0' and therefore we
stay in state \lq Odd'. The third input character is again a \lq 0'
so we remain in \lq Odd'. As fourth input we receive a \lq 1'
so we transition from \lq Odd' to \lq Even'. 
If we continue processing $100110001$ under this rationale,  the final outcome of the computation is the state \lq Even' as $100110001$ has even parity. Now, we use input $10\#010$. Characters \lq 0' and \lq 1' are processed as just described and,  since the third input character is \lq \#', $\cal{R}$ goes from state \lq Odd' to state \lq Error' and it remains there until the whole string is processed.

\subsubsection{Deterministic and nondeterministic computation}
\label{dndcomputation}

\begin{figure}
\begin{center}
\scalebox{0.4}{\includegraphics{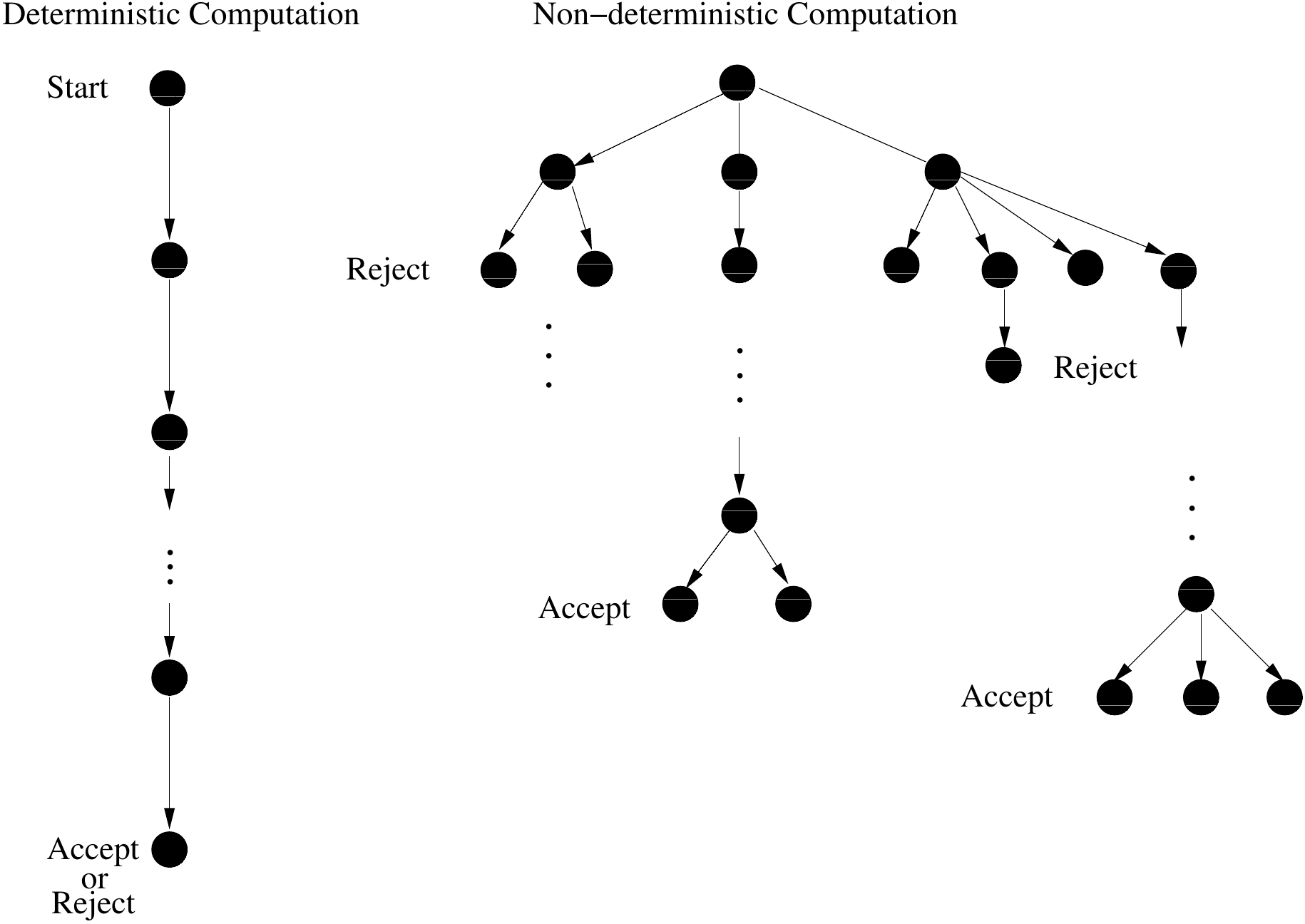}}
\end{center}
\caption{{\small  In deterministic computation, every single state of the computation is followed by only one state, given an input datum. In nondeterministic computation, a step may be followed by $n$ new steps.}}
\label{non_determinism}
\end{figure}


Each and every automaton has deterministic and nondeterministic versions (Fig.~(\ref{non_determinism})).  Deterministic automata follow a simple rule: given an input datum, every state of the computation is followed by only one state. Formally speaking, if $S$ is the set of states and $\Sigma$ is the set of input data, then any transition between states is defined by the function $\delta: S \times \Sigma \rightarrow S$ (depending on the mathematical properties of specific automata, we may add sets to the domain and codomain of $\delta$. Nevertheless, the key idea of this definition remains the same: given an input datum, every state of the computation is followed by only one state). In contrast, when computing on a nondeterministic machine, a state could be followed by {\it many} states, depending on the input datum and the transition rules.

Formally speaking, if $S$ is the set of states and $\Sigma$ is the set of input data then any transition between states is defined by the function $\delta: S \times \Sigma \rightarrow {\cal P}(S)$, where ${\cal P}(S)$ is the power set of $S$. In general, after reading an input symbol, a nondeterministic automata splits into multiple copies of itself and follows all the possibilities in parallel \cite{sipser06,savage98}.

Nondeterministic computation is a fictitious model of computation that can be thought of as parallel computation with unlimited resources. However, since resources are always limited, nondeterministic computation may seem to be an unreasonable model.  Despite this argument, there are two good reasons to use this model: i) to determine if a problem is computable even if  limitless resources are available, and ii) to define a key set of computable problems: $\mathsf{NP}$ problems (subsection \ref{computational_complexity}). Finally, please notice that, as stated in subsection \ref{pnpnpcsubsubsection}, i) $\mathsf{NP}$ problems have a fully realistic alternative definition in terms of polynomial time verifiers, and ii) both definitions are equivalent \cite{goldreich10}.

\subsubsection{Turing machines}
\label{turingmachines}


In \cite{turing36}, A. Turing introduced the famous abstract computing machines now known as {\it Turing Machines}. In addition to explaining the fundamental behaviour of modern computers, i.e. controlling the operation of a computer by means of a program stored in its memory, Turing presented the concept of {\it Universal Turing Machine} (UTM), that is, a Turing machine capable of  simulating any other Turing machine (the UTM would be analogous to a modern digital computer and the simulated Turing machine to a program encoded in the computer's memory). In this sense, Turing machines are thought of as the most general abstract computer model that can be formulated \cite{savage98}. 

We intuitively present Deterministic and Nondeterministic Turing machines (DTM and NTM, respectively) in order to introduce $\mathsf{P}$ and $\mathsf{NP}$ problems in subsection \ref{computational_complexity}. Turing machines are composed of states, transition rules, a language, a control unit and at least one limitless memory tape. Like any other automata:
\begin{itemize}
\item
Following the rationale presented in subsection \ref{dndcomputation}: 
\begin{itemize}
\item[--]
DTMs have a limited transition set. For every step of a computation on a DTM, a state is followed by only one state, i.e. $e_t \rightarrow e_{t+1}$ (in other words, state transition in a DTM is a function). 
\item[--]
NTMs have an unlimited amount of transitions. For every step of a computation on these machines, a state {\it may} be followed by one or more states, i.e  $e_t \rightarrow \{e^j_{t+1}\}$ (that is, state transition is a non-functional relation).
\end{itemize}

\item
Computation on a Turing machine starts in a pre-defined state and transitions among states depend on both transition rules and the input string fed to the Turing machine (switching among states is a job for the control unit). 
\item
The purpose of a Turing machine is to determine if an input string $\omega$ is an element of the language $L$ upon which the Turing machine has been defined. 
\begin{itemize}
\item[--]
{\bf Computation on a DTM}. If $\omega \in L$, then the DTM will end up in the accept state and we say that the DTM has accepted $\omega$; if  $\omega \notin L$ then the processing of $\omega$ will end up in the reject state of the DTM. For example, let $L$ be composed of strings $\omega$ made of 0s and 1s such that the last two characters of all $\omega$  are 00, and let $M$ be a DTM built to recognise $L$. Then, strings such as  $10100$ would be accepted by $M$ while $111$  would be rejected.
\item[--]
{\bf Computation on an NTM}. Nondeterministic computation produces many branches, as illustrated in Fig.~(\ref{non_determinism}). In general, the computation of any string on an NTM will produce branches ending in either accept or reject states. For a string $\omega$ to be accepted by an NTM, it suffices that {\it at least} one branch ends in an accept state.
\end{itemize}
\end{itemize}

\subsection{Computational complexity}
\label{computational_complexity}

Computational complexity focuses on two tasks: i) classifying problems in complexity classes, i.e. in terms of the amount of computational resources required to solve them, and ii) estimating the number of elementary steps required to run an algorithm. Complexity classes are abundant (\cite{complexityzoo16} comprises hundreds of them), being of utmost importance the $\mathsf{P}$, $\mathsf{NP}$, $\mathsf{NP-complete}$ and $\mathsf{NP-hard}$ classes \cite{mertens02,garey79,savage98,papadimitriou95,sipser06,goldreich10,moore11,AusielloBook}. 

\subsubsection{Tractable and intractable problems}

An algorithm is a step-by-step strategy defined to solve a problem in finite time $t \in \mathbb{N}$. To measure algorithm performance we use {\bf Asymptotic Analysis}, a technique in which we are interested in the maximum number of steps $S_{max}$ that an algorithm must run, given an input of size $n$.
We do so by considering only the highest order term of the expression that quantifies $S_{max}$. For example, the highest order term of $S_{max}(n) = 18n^9 + 8n^5 - 3n^4$  is $18n^9$ and, since we disregard constant factors, we say that $S_{max}$ is asymptotically at most $n^9$. The following definition formalises this idea.

\begin{definition}{\bf Big $O$  notation}. Let $f,g: \mathbb{N} \rightarrow \mathbb{N}$.  We say that $f(n) = O(g(n))$ iff $\exists$ $\alpha,n_o \in \mathbb{N}$ such that $\forall \text{ } n \geq n_o$ $\Rightarrow$ $f(n) \leq \alpha g(n)$.
 Then, $g(n)$ is an asymptotic upper bound for $f(n)$  and we say that $f$ is of the order of $g$. Informally, $f(n) = O(g(n))$ means that $f$ grows as $g$ or slower. 
\end{definition}

The time complexity of an algorithm $A$  estimates, for each input length, the largest amount of time needed by $A$ to solve any problem instance of that size (that is, this is the worst-case over all inputs of size $n$). Algorithms with time complexity function $O(p(n))$, where $p(n)$ is a polynomial and $n$ is  the input length,  are  {\it polynomial-time} algorithms. Algorithms whose time complexity function cannot be so bounded are called an {\it exponential-time} or {\it factorial-time} algorithms.

\begin{figure}[hbt]
\hfill{}
\epsfig{width=4.2in,file=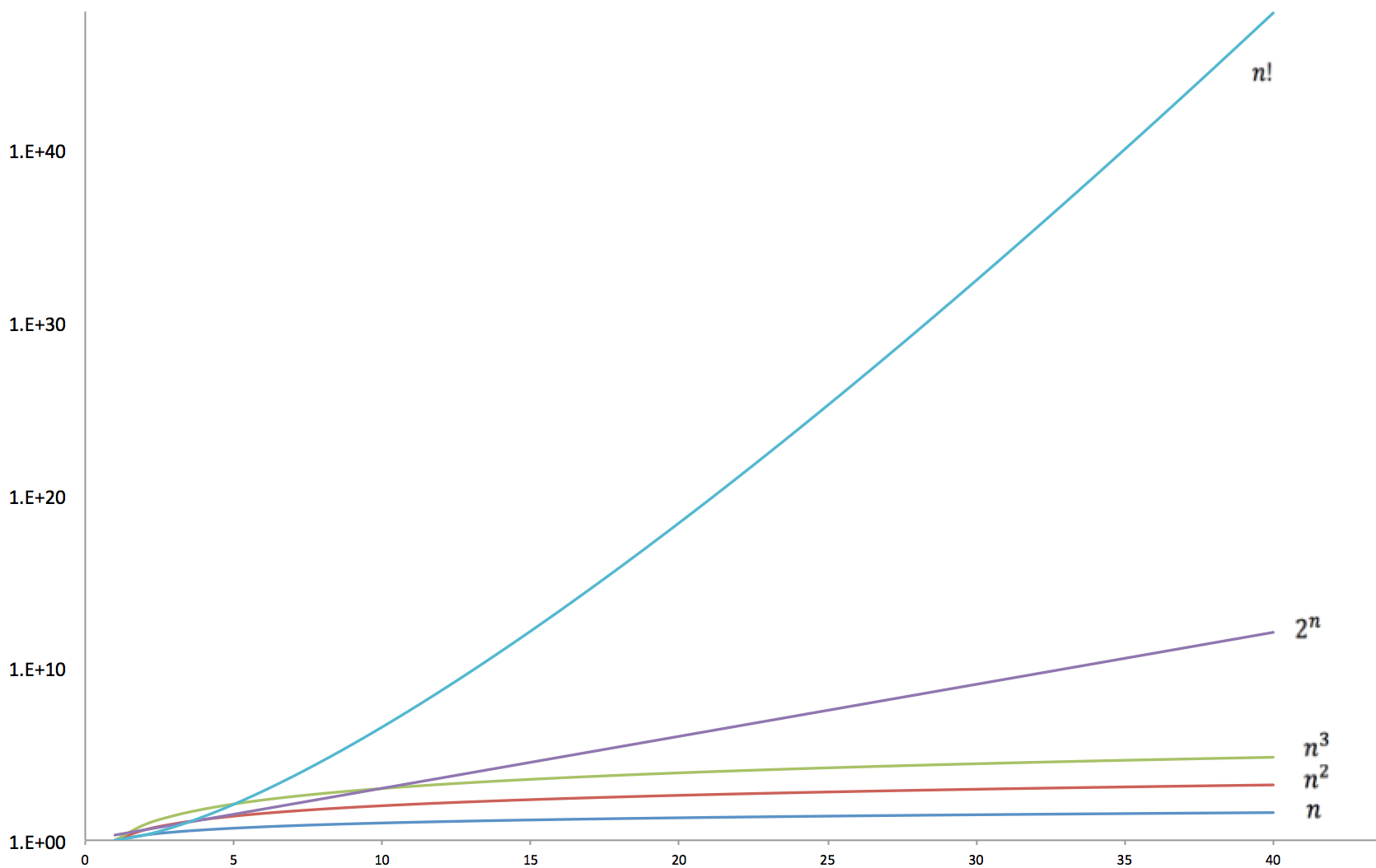}
\hfill{}
\caption{{\small Time complexity functions  $T_{A_1} = n$, $T_{A_2} =n^2$, $T_{A_3} =n^3$, $T_{A_4} =2^n$, and $T_{A_5} =n!$. X axis values are the length of input strings for algorithms $A_i$ while values on the Y axis are time steps in logarithmic scale. Note that even for modest input lengths, exponential and factorial functions consume massive amounts of time steps (for instance, if 1 time step = 1s, then $T_{A_5}(50) \approx 3.17\text{E+52}$ years).}
}\label{complexity_graph}
\end{figure}

For example, let us suppose we have algorithms $A_1$, $A_2$, $A_3$, $A_4$ and $A_5$ with corresponding time complexity functions given by $n$, $n^2$, $n^3$, $2^n$, and $n!$ as shown in Fig.~(\ref{complexity_graph}). Note that time complexity functions $T_{A_1} = n, T_{A_2} =n^2, T_{A_3} =n^3$ grow much slower than $T_{A_4} =2^n$ and, even worse, than $T_{A_5} =n!$. Since i) polynomials grow strictly slower than exponential functions  \cite{papadimitriou95},  and ii) $n! > 2^n$ for all $n\geq 4$, we can see why polynomial-time algorithms are considered as acceptable  (because their resource consumption is moderate when compared to exponential/factorial-time algorithms). In contrast, exponential- and factorial-time algorithms can easily consume massive amounts of time steps even for relatively small input sizes, which is why those types of algorithms are not considered satisfactory solutions. Problems that are solved by polynomial-time algorithms are known as {\it tractable} problems while problems that are solved by exponential/factorial-time algorithms are known as {\it intractable} problems.


\subsubsection{$\mathsf{P}$, $\mathsf{NP}$, $\mathsf{NP-complete}$ and $\mathsf{NP-hard}$ problems}\label{pnpnpcsubsubsection}

Decision problems, a key concept in complexity theory,  can be informally defined as questions with yes/no answers. Formally, a decision problem can be defined as follows:

\begin{definition}\cite{savage98}. {\bf Decision problem}. Let $\Sigma$ be an alphabet and $\Sigma^*$ a language over $\Sigma$.  A decision problem  $\mathcal{D}$ is a set of instances $I \subseteq \Sigma^*$ of $\mathcal{D}$ and a condition $\phi_{\mathcal{D}}: I \longmapsto \{ 0,1\}$ that has the value $1$ on \lq Yes' instances and $0$ on \lq No' instances. 
\label{decisionproblem}
\end{definition}

Decision problems can be put in correspondence with languages as follows: given an encoding scheme, an arbitrary instance $s$ of a decision problem $\mathcal{D}$ can be transformed into a member of a language $L_1$, i.e. a string  $t \in L_1$.  Moreover,  we can take $t$ as input of a Turing machine $M$ defined upon another language $L_2$ and, if $M$ accepts (rejects) $t$, we say that the answer for $s$ is Yes (No).  For instance,  determining if an integer number is divisible by four can be formulated as the following decision problem: {\it for a given number $n$, is $n \mod 4 = 0$?} This decision problem can be solved by rewriting $n$ as a binary number and using the Turing machine presented  in subsection \ref{turingmachines} (when expressed as binary numbers, integer numbers divisible by 4 are strings ending in $00$). Let us now define complexity classes $\mathsf{P}$ and $\mathsf{NP}$.

\begin{definition}\cite{savage98}. {\bf $\mathsf{P}$ and $\mathsf{NP}$ problems}. 
The classes  $\mathsf{P}$ and $\mathsf{NP}$ are sets of decision problems solvable in polynomial time on DTMs and NTMs respectively. So,
\\

-- For each decision problem $\mathcal{A} \in  \mathsf{P}$  there is a DTM $M$ and a polynomial $p(n)$ such that i) processing input  $\omega$ of length $n$ in $M$ would be done in $p(n)$ steps at most, and ii) $M$ would stop in the accept (reject) state iff $\omega$ belongs (does not belong) to $\mathcal{A}$. 


--For each decision problem $\mathcal{B} \in  \mathsf{NP}$  there is an NTM $N$ and a polynomial $p(n)$ such that i) processing input $\omega$ would be accepted by $N$ iff at least one computation branch ends in an accept state, and ii) an accepting computation for input  $\omega$ of length $n$  requires $p(n)$ steps.

-- Please bear in mind that a computation state on a DTM is a function, i.e. a state is followed by only one state ($e_t \rightarrow e_{t+1}$) while a computation step on an NTM is a one-to-many relationship, i.e.  a state {\it may} be followed by one or more states ($e_t \rightarrow \{e^j_{t+1}\}$).

-- The relationship between algorithmic time complexity and solving a decision problem using a Turing machine is: algorithms are translated into transition rules and each algorithmic step must be performed in a polynomial number of Turing machine steps.

\label{pandnp}
\end{definition}

We may also define  $\mathsf{NP}$ as the set of decision problems for which it is possible {\bf to check}, in polynomial time, if a proposed solution is indeed a solution. Formally, $\mathsf{NP}$ is the class of decision problems that have polynomial time verifiers \cite{sipser06} and it is possible to prove that both definitions presented in this section are equivalent (theorem (2.8) of \cite{goldreich10}). For example,  let $\mathcal{D}$ be the problem of finding the prime factors of an integer. It is possible to quickly check if a proposed set of integers is indeed the prime decomposition of a given integer (e.g. the numbers $\{5,7,13,17,29\}$ are the prime factors of $224315$).

NTMs and DTMs solve the same set of decision problems \cite{sipser06}, i.e. any NTM can be simulated by a DTM, being the only remarkable difference the amount of resources needed to do so. Theorem~(\ref{pandnpexponentialgap02}) establishes an upper bound relationship for resource consumption in deterministic and nondeterministic Turing machines.

\newpage{}

\begin{theorem} \cite{garey79}. If decision problem $\cal{A} \in \mathsf{NP}$ then there is a polynomial $p(n)$, where $n$ is the size of an arbitrary input data, such that $\cal{A}$ can be solved by a deterministic algorithm having time complexity $O(2^{p(n)})$.
\label{pandnpexponentialgap02}
\end{theorem}

 The question ${\mathsf P} \stackrel{?}{=} \mathsf{NP}$ is a fundamental problem in theoretical computer science (it is known that $\mathsf{P} \subseteq \mathsf{NP}$ and it is unknown whether $\mathsf{NP} \subseteq \mathsf{P}$). A positive answer would mean that {\it finding solutions to decision problems is as hard (or as easy) as checking the correctness of solutions} \cite{goldreich10}. Equivalently,  ${\mathsf P} = \mathsf{NP}$  would mean that finding efficient algorithms for any $\mathsf{NP}$ problem is just a matter of ingenuity and hard work, i.e. there would be no fundamental reason to suppose that any $\mathsf{NP}$ problem {\bf must} be intractable. A key research area in computer science is the study of $\mathsf{NP}$ problems  for which we only know algorithms with time complexity functions that are not upper-bounded by polynomials. An example of that kind of  $\mathsf{NP}$ problems are the  $\mathsf{NP-complete}$ problems.

Albeit closely related, the mathematical definitions of $\mathsf{NP-hard}$ and $\mathsf{NP-complete}$ problems are different. In \cite{knuth74}, D. Knuth introduces the following layperson definitions: \lq \lq $\mathsf{NP-hard}$ means as hard as the most difficult problem in NP, while $\mathsf{NP-complete}$ means representative of the complete class $\mathsf{NP}$ with respect to difficulty''. Formal definitions of $\mathsf{NP-complete}$ and $\mathsf{NP-hard}$ problems follow:

\begin{definition}{\bf Karp-reduction}. Let $\cal{A},\cal{B}$ be two decision problems and $f$ a function that can be computed in polynomial-time. Then, $f$ is a Karp-reduction of $\cal{A}$ to $\cal{B}$ iff, for every instance $x$ of $\cal{A}$ we find that   $f(x)$ is an instance of $\cal{B}$.
\end{definition}

\begin{definition}{\bf $\mathsf{NP-complete}$ problems} \cite{goldreich10}. Let $\cal{X}$ be a decision problem. $\cal{X}$ is an $\mathsf{NP-complete}$ problem if and only if the following two conditions hold:
\\
i) $\cal{X} \in \mathsf{NP}$.
\\ 
ii) For each problem ${\cal B}_i \in \mathsf{NP}$ there is a Karp-reduction $f_i : {\cal B}_i  \rightarrow \cal{X}$. 
\label{npc}
\end{definition}

\begin{definition}{\bf $\mathsf{NP-hard}$ problems} \cite{goldreich10}. A problem $\cal{Y}$ is an $\mathsf{NP-hard}$ problem if and only if every problem ${\cal B}_i \in \mathsf{NP}$ is Karp-reducible to $\cal{Y}$. {\it Note that $\mathsf{NP-hard}$ problems are not {\it required} to be $\mathsf{NP}$ problems}.
\label{nph}
\end{definition}


We now introduce the K-SAT problem, a key  decision problem in computer science.

\begin{definition}{\bf The K-SAT Problem} \cite{sipser06}. Let $S=\{x_1, x_2, \ldots, x_n, \bar{x}_1, \bar{x}_2 \ldots \bar{x}_n \}$ be a set of Boolean variables and their negations. We define a clause $c$ as a disjunction of binary variables in $S$ (for example, $c = x_3 \vee \bar{x}_8 \vee \bar{x}_{13}$). We now define $\Phi$ as a conjunction of clauses $c_i$ over $S$ where each clause $c_i$ has $K$ variables, i.e. $\Phi$ is a conjunction of disjunctions $$\Phi = \bigwedge_i c_i = \bigwedge_i [(\bigvee_{j=1}^K \overset{(-)}{x_{\alpha_j}}],$$ where $\alpha_j \in \{1, 2, \ldots n\}$ and $\overset{(-)}{x_{\alpha_j}}$ is used to denote either $x_{\alpha_j}$ or $\bar{x}_{\alpha_j}$. A logical formula written as a conjuction of disjunctions, such as $\Phi$, is said to be written in {\bf conjunctive normal form}. 

The K-SAT problem consists of determining whether a Boolean expression such as $\Phi$ is satisfiable or not, i.e. whether there is a set of values of $\{x_1, x_2, \ldots, x_n \}$ for which $\Phi=1$. In addition to its most important role in theoretical computer science \cite{garey79,sipser06}, the K-SAT has several key applications in many branches of science and engineering like artificial intelligence and manufacturing \cite{marquessilva08}.

\label{3sat}
\end{definition}
For example, let $\{x_1, x_2, x_3, x_4, x_5, x_6, x_7, x_8, x_9 \}$ be a set of binary values and
\begin{eqnarray}
\Phi &=&
(\overline{x}_3 \vee x_6 \vee x_9 ) \wedge
(x_2 \vee \overline{x}_4 \vee x_6 ) \wedge
(\overline{x}_1 \vee \overline{x}_7 \vee \overline{x}_8 )\wedge
(\overline{x}_3 \vee x_4 \vee x_9 ) \wedge \nonumber\\
&& (\overline{x}_1 \vee \overline{x}_5 \vee \overline{x}_8 )\wedge
(x_4 \vee \overline{x}_5 \vee \overline{x}_8 )\wedge
(x_2 \vee x_3 \vee x_5 ) \wedge
(\overline{x}_4 \vee \overline{x}_5 \vee x_7 ) \wedge \nonumber\\
&& (\overline{x}_2 \vee x_5 \vee x_6 ) \wedge
(x_1 \vee x_3 \vee \overline{x}_4 )\wedge
(\overline{x}_1 \vee x_5 \vee \overline{x}_9 )\wedge
(x_1 \vee x_3 \vee x_4 ) \wedge \nonumber\\
&& (x_3 \vee \overline{x}_6 \vee \overline{x}_7 )\wedge
(x_4 \vee x_6 \vee \overline{x}_7 )\wedge
(\overline{x}_4 \vee \overline{x}_7 \vee x_8 ) \wedge
(\overline{x}_1 \vee x_2 \vee \overline{x}_4 )\wedge \nonumber\\
&& (x_1 \vee \overline{x}_4 \vee x_7 ) \wedge
(x_6 \vee x_8 \vee x_9 ) \wedge
(x_4 \vee \overline{x}_5 \vee \overline{x}_9 )\wedge
(x_2 \vee \overline{x}_4 \vee x_8 ) \wedge \nonumber\\
&& (\overline{x}_2 \vee \overline{x}_3 \vee \overline{x}_8 )\wedge
(x_1 \vee \overline{x}_5 \vee x_7 ) \wedge
(\overline{x}_2 \vee \overline{x}_3 \vee \overline{x}_5 )\wedge
(\overline{x}_4 \vee x_5 \vee x_7 ) \wedge \nonumber\\
&& (\overline{x}_2 \vee \overline{x}_6 \vee \overline{x}_8 )\wedge
(\overline{x}_4 \vee \overline{x}_6 \vee x_9 ) \wedge
(x_7 \vee \overline{x}_8 \vee \overline{x}_9 )\wedge
(x_2 \vee \overline{x}_3 \vee x_5 ) \wedge \nonumber\\
&& (\overline{x}_2 \vee \overline{x}_4 \vee x_5 ) \wedge
(\overline{x}_1 \vee \overline{x}_4 \vee \overline{x}_6 )\wedge
(x_1 \vee x_4 \vee x_5 ) \wedge
(x_5 \vee \overline{x}_7 \vee \overline{x}_9 )\wedge \nonumber\\
&& (x_1 \vee x_2 \vee x_4 ) \wedge
(\overline{x}_2 \vee x_3 \vee \overline{x}_6 )\wedge
(\overline{x}_4 \vee x_7 \vee \overline{x}_8 )\wedge
(x_1 \vee \overline{x}_3 \vee x_6 ) \wedge \nonumber\\
&& (\overline{x}_3 \vee \overline{x}_4 \vee x_5 ) \wedge
(\overline{x}_1 \vee \overline{x}_3 \vee x_7 ) \wedge
(\overline{x}_1 \vee x_3 \vee x_9 ) 
\label{example3sat}
\end{eqnarray}

be an instance of 3-SAT (i.e. K-SAT with K=3).  Finding the solutions (if any) of even a modest 3-SAT instance like Eq. (\ref{example3sat}) can become difficult quite easily ($\Phi$'s only solution is $x_1 = 0, x_2 = 0, x_3 = 1, x_4 = 1, x_5 = 1, x_6 = 1, x_7 = 1, x_8 = 1, x_9 = 1)$.
\\

Two key properties of K-SAT are:
\\
\\
- K-SAT is an $\mathsf{NP}$ problem. This follows from the fact that solving any instance $\Phi$ of the K-SAT problem requires i) the computation of $2^n$ different combinations of binary values for $\{x_i\}$, and ii) substituting each binary string (from $00 \dots 0$ to $11 \dots 1$) in $\Phi$. Assuming an exponential amount of resources (i.e., an NTM), this brute-force algorithm runs in polynomial time. Moreover, the correctness of any solution, i.e. any proposed set of values for the binary values $\{x_1, x_2, \ldots, x_n \}$ for an arbitrary instance $\Phi$ of K-SAT, can be checked in polynomial time (we just substitute the $n$ values on $\Phi$, which is composed of a polynomial number of variables and logical operations).  
\\
- K-SAT are $\mathsf{NP-complete}$ for all K $\ge 3$ \cite{cook71,levin73,garey79}.
\\

Optimisation can be defined as the process of finding optimal solutions \cite{du16}. Given a function $f:A \mapsto B$,  the optimisation problem consists of finding the value(s) $x$ within the domain of $f$ such that $f(x)$ is an extremum, i.e. a maximum or minimum value in the range (or a subset of the range) of $f$, possibly subject to a number of constraints \cite{press92}. Global optimisation, i.e. the process of finding the highest or lowest value  within the full range of a function $f$, is indeed a most difficult and challenging problem in science and engineering. 

Optimisation can be classified into discrete or continuous, depending on whether variables belong to discrete or dense sets.

Known exact algorithms for solving discrete optimisation problems are brute-force algorithms, i.e. computational procedures that try every possible solution. Running exact algorithms for large instances of discrete optimisation problems on classical digital computers is unreasonable because of the colossal amount of time or computer resources such algorithms would require. A practical solution to this dilemma is provided by approximation algorithms as they are capable of producing approximate solutions reasonably fast.  It is pretty much a given in research on combinatorial optimization problems (Def. \ref{combopdef}) that, in practical terms, we have no choice but to be happy with a non-optimal solution,  and generally have no way of knowing how close it might be to optimal.  As for continuous optimisation problems, approximate solutions are compulsory because of floating-point representation of real numbers on classical digital computers.

\newpage{}

Optimisation techniques are generally divided into derivative or non-derivative methods, depending on whether the computation of derivatives of the objective function is required or not \cite{du16}. Examples of derivative techniques are the 1D Golden Section Search and variable metric methods like Fletcher-Powell and BFGS techniques \cite{press92}. Examples of non-derivative methods are abundant and can be found to solve both discrete and continuous optimisation problems. In the following lines we shall mention some examples of non-derivative methods.

Broadly speaking, optimisation algorithms can be classified according to the following categories: analytical, enumeration and heuristic search techniques \cite{du16}. Analytical algorithms are built upon mathematical procedures designed to look for exact or $\epsilon$-close solutions and tend to heavily use calculus-based operators and procedures, enumeration algorithms aim at comprehensively exploring the search space of an objective function, and heuristic search algorithms are problem-generic solution strategies, sometimes inspired on physical phenomena or the behaviour of natural entities, that may provide sufficiently good solutions to an optimisation problem. Heuristic search is of great importance in computer science, algorithmics and many branches of industry because they can often find approximate solutions to key problems while consuming less computational resources (e.g., in terms of time complexity functions) than analytical and enumeration algorithms \cite{du16,press92,guenin15,mcgeoch-qa}. Examples of heuristic algorithms are Particle Swarm Optimisation and Simulated Annealing.


Optimisation problems are not in  $\mathsf{NP}$ because they are not decision problems. 
Also, due to resource consumption, solving an optimisation problem can be as hard as solving an $\mathsf{NP}$ problem. So, optimisation problems are examples of $\mathsf{NP-hard}$ problems. 



The SAT problem consists of determining if there is a set of Boolean variables that satisfies a given Boolean formula \cite{moore11}. Since every Boolean formula can be transformed into an equivalent Boolean expression expressed in  conjunctive normal form, we can see that every instance of the K-SAT problem has a corresponding SAT instance. The maximum satisfiability (max-SAT) is an optimisation version of SAT and it consists of finding a truth assignment that satisfies as many clauses of a given instance as possible. The max-SAT problem is an example of a combinatorial optimisation problem. We shall explore how to solve instances of the max-SAT problems with quantum annealing algorithms in section \ref{william}.

\begin{definition}{\bf Combinatorial Optimisation Problem} \cite{press92,neosguide}. Let $E$ be a finite set with cardinality $|E| = n$, $P_E$ the power set of $E$ (hence $|P_E| = 2^n$) and the function $C: P_E \rightarrow \mathbb{R}$. The general setup of a combinatorial optimisation problem is to find an element $\mathcal{P} \in P_E$ such that $C(\mathcal{P}) = min_{\mathcal{P}_i \in P_E} \{C(\mathcal{P}_i)\}$. The term combinatorial comes from $\sum_{k=0}^{n} {n \choose k} = 2^n$. The discrete domain of combinatorial optimisation problems make them suitable for algorithmic analysis, as we shall see in the following sections.
\label{combopdef}
\end{definition}

Assuming ${\mathsf P} \neq \mathsf{NP}$, no optimisation algorithm for an $\mathsf{NP-hard}$ problem can be both {\em fast} (running in polynomial time) and {\em optimal} (guaranteeing optimal solutions) on {\em all} inputs.  However, many heuristic algorithms have been proposed that can meet two out of three of these objectives:  for example one might be always optimal but only fast on some inputs, and another might be always fast but sometimes non-optimal (returning good-to-poor solutions), depending on the input.  Simulated annealing and quantum annealing, discussed in the following sections, are heuristics of this type, with parameters that allow the programmer to trade speed for solution quality.  

\subsection{Simulated and quantum annealing}

{\bf Simulated annealing}. The Simulated Annealing algorithm (SA) \cite{kirkpatrick83} is an example of a heuristic, i.e. an algorithm capable of finding good solutions reasonably quickly on many types of inputs. SA has several parameters that can be modified to trade speed for solution quality.


\newpage{}
  
The experimental technique upon which SA is based is that of finding the equilibrium state of a system in the limit of low temperature. In this regard, the process identified to produce robust experimental results is known as {\it annealing} and it consists of three steps: a) melting the material under study, b) lowering the temperature very slowly, and c) spending a considerable amount of time in a range of temperatures close to freezing point \cite{kirkpatrick83}.

If we rephrase this method in mathematical parlance, we may say that the purpose of this process is to find the state of a system that corresponds to the minimum value of a variable (energy) and we want to do it by manipulating another variable (temperature). In other words, we have a physical process, {\it annealing}, that resembles the behaviour of an optimisation algorithm. This is the motivation for developing SA.

The set of instructions (i.e., the pseudocode) of SA is presented in Algorithm (\ref{algorithm_sa}) \cite{henderson03}.  Suppose that we have a combinatorial optimisation problem $\Lambda(\alpha_1, \alpha_2, \ldots, \alpha_n)$ where each $\alpha_i$ is a binary variable and ${\cal D}$ is the set of all possible combinations of $\alpha_i$. Also, for each element $\boldsymbol \alpha \in {\cal D}$ we define $N(\boldsymbol \alpha)$, the neighborhoord of $\boldsymbol \alpha$, which consists of elements $\boldsymbol \alpha_i \in {\cal D}$ that are close to $\boldsymbol \alpha$, according to a given rule or metric. Finally,  let $f:\cal{D} \mapsto \mathbb{R}$ be a cost function that assigns a value to each element of $\cal{D}$. Our aim is to minimise $f$.

\begin{center}
\begin{algorithm}[H]
 \label{algorithm_sa}
 
 Select an initial value $\omega \in {\cal D}$\;
 Select the temperature change counter $k=0$\;
 Select a temperature cooling schedule, $t_k$\;
 Select an initial temperature $T = t_0 \ge 0$\;
Select a repetition schedule, $M_k$, that defines the number of iterations executed at each temperature $t_k$\;
 \Repeat{stopping criterion is met}{
   Set repetition counter m = 0\;
   \Repeat{$m=M_k$}{
   Generate a solution $\omega' \in N(\omega)$\;
   Calculate $\Delta_{\omega,\omega'} =  f(\omega') - f(\omega)$\;
   If  $\Delta_{\omega,\omega'} \le 0$ then $\omega := \omega'$\;
   If  $\Delta_{\omega,\omega'} > 0$ then $\omega := \omega'$ with probability $exp(-\Delta_{\omega,\omega'}/t_k)$\;
   $m:= m+1$\;
    }
    $k:= k+1$\;
    }
 \caption{Simulated annealing pseudocode}
\end{algorithm}
\end{center}

SA is a powerful heuristic algorithm that has been used to approximately solve many hard problems, among them the most famous Traveling Salesman Problem \cite{press92}.
\\

{\bf Quantum annealing}. Building quantum algorithms is a most challenging task because of two reasons: a) quantum mechanics is a counterintuitive theory and intuition plays a key role in algorithm design, and b) a competitive quantum algorithm must not only produce the solutions it is expected to, it also has to be more efficient, at least for some input values, than any classical algorithm (at least more efficient than existing classical algorithms). 



Quantum annealing (QA) \cite{kadowaki98} is a heuristic inspired in SA that can be used to solve optimisation problems. Instead of using thermal fluctuations as SA does, QA uses quantum tunneling to move through the landscape defined by the cost function associated to the optimisation problem at hand. 

Let us briefly describe the rationale behind quantum annealing. A Hamiltonian designed for quantum annealing can be written as 

\begin{equation}
{\cal H} = {\cal H}_F + \Gamma(t){\cal H}_D
\label{qaintro01}
\end{equation}

where ${\cal H}_F$  is a Hamiltonian that encodes the function to be optimised, ${\cal H}_D$ is another Hamiltonian that introduces an external  transverse field and $\Gamma$ is a transverse field coefficient that is used to control the intensity of the external field \cite{mcgeoch-qa}.

For example, let us introduce the following Hamiltonian

$$
{\cal H}_{example} = - \sum_{\langle i,j \rangle} J_{i,j} \sigma_i^z \sigma_j^z - \Gamma (t) \sum_i \sigma_i^x
$$

where $\sigma^z$ $\sigma^x$ are Pauli operators. The physical effect of $\Gamma (t)$ is to induce flips of $\sigma_z$. Note that $\Gamma$ is a function of time, hence the intensity of the transverse field will change in time. Typically for a quantum annealing algorithm, $\Gamma$ will be intense at the beginning and it will gradually decrease as the algorithm is executed (hence, intense quantum tunnelling is expected at the beginning and it will decrease as the algorithm runs) \cite{battaglia06}.

Conceptually speaking, the rationale behind quantum annealing is to run a physical process that evolves according to a Hamiltonian described by Eq.(\ref{qaintro01}). If the system evolves very slowly (i.e adiabatically), it will eventually settle in a final ground state that, with certain probability, will correspond to the optimal value of the function encoded in ${\cal H}_F$. Exact simulation of a quantum annealing process on a digital computer is costly \cite{bapst11}.
\\

{\bf Simulated quantum annealing}.  Quantum Monte Carlo methods are stochastic procedures that are used to solve the Schr\"odinger equation. They consist of using classical Markov Chain Monte Carlo methods to estimate low-energy states of Hamiltonians \cite{battaglia06,crosson2017}.

Simulated quantum annealing (SQA) is a classical algorithm that uses Quantum Monte Carlo methods to simulate quantum annealing Hamiltonians \cite{crosson2017}. The general performance of SQA (i.e. SQA's computational complexity) is still unknown. However, some encouraging results can be found in \cite{martonak02,battaglia06} where Path-Integral Quantum Monte Carlos methods are shown to be very effective for simulating some QA Hamiltonians as well as in \cite{crosson2017} where it is proved that SQA can be exponentially faster than SA for optimising the {\it Hamming weight with a spike function}, which is a toy model used to contrast the computational power of QA and SQA.

The setup of QA-based algorithms and their implementation on D-Wave quantum annealers will be explored in detail in the following sections.


\section{D-Wave's annealing-based quantum computers} 
\label{dwavecathy}

D-Wave Systems manufactures computer systems that implement a quantum annealing algorithm in hardware.    
Within the system a {\em quantum processing unit} (QPU) operates on qubits (quantum bits) that behave as 
particles in a physical quantum process, exploiting properties such as superposition and entanglement,  
to find low-energy states that correspond to low-cost solutions to optimisation problems.

The motivation for developing these systems comes from theoretical arguments that quantum computation 
could be much faster than classical computation at some tasks.  This does not mean that quantum machine instructions  
are faster than classical ones;  rather, the conjecture is that for some problems, quantum algorithms might run 
faster by ``skipping steps''  that would be necessary to classical algorithms.   Quantum algorithms operate on qubits  that have special properties not available to bits in classical computers. Two important properties are: {\bf i) Superposition}. Whereas a classical bit has value either 0 or 1, a quantum bit can be in a combination of both 0 and 1 at once. Like Schr\"odinger's cat, qubits experience superposition while isolated from the ambient environment: when a qubit is observed, it probabilistically \lq \lq collapses'' to either 0 or 1. {\bf ii) Entanglement}. A collection of qubit superposition states can become correlated in such a way that multiple qubits can change state simultaneously.


In the context of annealing methods, these properties mean that the quantum algorithm can exploit new ways to explore a given solution landscape, which are not available to classical algorithms. For example, simulated annealing  uses an array of $N$ bits to represent the current state of the solution, and the main calculation involves moving from state to state by changing one bit at a time. In contrast, because of superposition, a quantum algorithm can use $N$ qubits to probabilistically represent {\it all} of the solution space at once; and because of entanglement it can modify groups of superposition probabilities in constant time rather than qubit-by-qubit. Under certain conditions, these two properties can lead to {\it tunnelling}, which allows the quantum algorithm to pass through hills in the solution landscape instead of climbing over in the classical way.

Quantum phenomena like quantum tunneling and quantum entanglement do not have known classical analogs. The simulation of such quantum phenomena in classical digital computers is made by manipulating classical hardware via software (i.e., by programming the mathematical description of quantum phenomena) but the computational resources needed to do so (e.g. memory or execution time) rapidly scale hence making this kind of simulation unfeasible (e.g. \cite{galvao03}). In contrast, quantum hardware would be capable of exact quantum simulation (for example, \cite{sornborger12}). For example, in \cite{denchev} it has been shown that, for a crafted problem with tall and narrow energy barriers separating local minima, finite-range tunneling on a D-Wave 2X quantum annealer would present runtime advantages with respect to simulated annealing.

These capabilities suggest that potentially enormous speedups over classical computation times may be possible on some classes of input, but this potential must be set against certain challenges associated with controlling qubit states and transitions, which tend to erode the probability of successful outcomes. Understanding how these two factors balance out in D-Wave QPUs is a vigorous area of current research.

The class of $\mathsf{NP-hard}$ problems described in section \ref{computerscience} represents a promising set of candidates for realising this potential.  Note that although optimisation is a natural fit to quantum annealing,  the algorithm has wider applications such as sampling and counting solutions and simulating quantum processes.  For example,  Machine Learning is an important application area that requires batches of  near-optimal solutions sampled according to specific probability distributions.    
%
%


The quantum annealing algorithm implemented in a D-Wave QPU is designed to be fast (in terms of absolute runtime) and to return optimal or near-optimal solutions;  some parameters are available to adjust this trade-off.  

The remainder of this section gives an overview of a D-Wave system from the user's perspective,  and briefly surveys the technology stack.  Some of this material is adapted from \cite{mcgeoch-qa} (Chapter 4).     

\begin{figure}
\begin{centering} 
\includegraphics[height=2.8in]{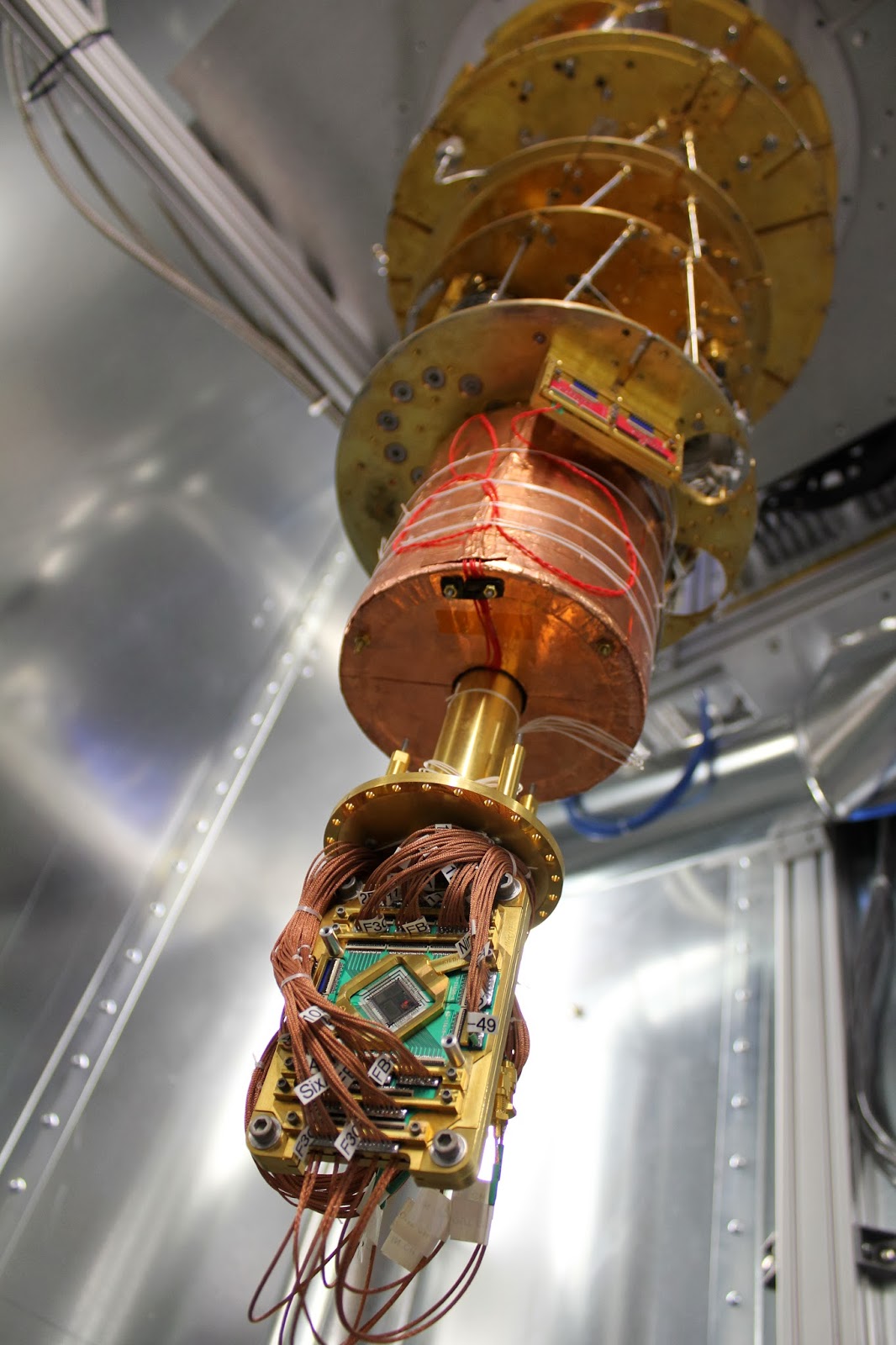} \\
\end{centering} 
\caption{The QPU chip is the black diamond at the bottom of the mounting apparatus shown here.  The bundled wires 
carry control and I/O signals between the QPU and the control system. }  
\label{mounting} 
\end{figure}

\subsection{The User's View} 
A D-Wave system comprises two main components:
\begin{enumerate} 
\item The QPU chip that implements the quantum annealing algorithm.  Fig.~(\ref{mounting}) shows a photograph of a QPU in 
its mounting apparatus.  

\item A conventional computer containing a front end server called the {\em solver application programming interface} (SAPI),  and a back end system that communicates with the QPU.  
\end{enumerate}  

\newpage{}

Given an input instance I for an $\mathsf{NP-hard}$ optimisation problem, there are two approaches to solving it using a D-Wave quantum annealing processor.  The first is to translate it to an input for an Ising model (IM) problem and then to transform it to a so-called {\em native} problem that matches the qubit connection topology of the D-Wave processor:  these steps are discussed in detail in section 4.   The second approach, not addressed here, is to solve the problem as-is using problem decomposition and a hybrid classical-plus-quantum-query approach; see \cite{mcgeoch-wang} or \cite{mwesigwa} for examples  of hybrid uses. 
\\

{\bf Transformation to Ising model}. Inputs for a general problem in $\mathsf{NP}$ must be transformed to inputs for IM.  This is done using standard techniques of NP-completeness theory (see section \ref{william} or \cite{lucas-formulations} for examples).   
The Ising Model problem is defined as follows:  given a graph $G=(V,E)$ with weights $h_i$ (called {\em fields}) on vertices and $J_{ij}$
(called {\em couplers}) on edges,  find an assignment of {\em spins} $S = (s_1, \ldots s_n)$ to vertices, with $s_i \in \{ -1, +1\}$, so as to minimise the {\em energy function} 
\begin{eqnarray}
E(S)  = \sum_{i \in V} h_i s_i   +  \sum_{(i,j) \in E} J_{ij} s_i s_j. 
\label{energyfunction}
\end{eqnarray}



{\bf Mapping $G$ to the Chimera topology}. An IM problem defined on a general graph $G$ must be translated to an equivalent problem that matches the connectivity structure of the QPU.  In current QPU models, qubits and couplers are structured according to a {\em Chimera graph}, described in the next section. The current fabrication process leaves a small percentage of qubits and couplers inoperable;  the {\em working graph} for any particular QPU is a subgraph $H \subset C$ of a Chimera graph.   

Translating a problem on $G$ to an identical problem on $H$ involves a process called {\em minor embedding}, discussed further in section \ref{william}.  
SAPI provides utilities for minor embedding,  so this process need not be carried out by hand.  However,  
more compact embeddings can sometimes by found by a custom approach,  especially when $G$ contains regular or local structures.  See  \cite{bian-csp,trummer,venturelli-jobshop} for more examples of 
problem-specific embeddings.  

The result of minor-embedding is to produce a {native} input  $I = (h_i, J_{ij})$ defined on the working graph $H$. Components  of $H$ that do not appear in $I$ have weight zero.  This input is ready to be sent to SAPI for solution on the QPU,  together with system parameters specifying how the computation should go.   One set of parameters describes the quantum annealing algorithm;  another set describes preprocessing and  post-processing strategies to improve results.\\


{\bf The quantum annealing algorithm}. 



D-Wave's QPU implements an Ising spin glass model on $n$ particles described by the Hamiltonian

\begin{eqnarray}
 {\cal H}_p   & = &  \sum_i  h_i \sigma^z_i  +  \sum_{i < j}  J_{ij} \sigma^z_i  \sigma^z_j
\end{eqnarray} 

where $\sigma^z_i$ is the Pauli matrix $z$ acting on particle $i$,  $h_i$  is the magnetic field on particle $i$,
and $J_{ij}$ is the coupling strength between particles $i$ and $j$.  The  minimum energy state, also called the ground state,  of ${\cal H}_p$ corresponds to a spin configuration $S=(s_1,\dots,s_n)\in\{+1,-1\}^n$ that minimises the Ising energy function in Eq. (\ref{energyfunction}).

Quantum annealing uses an analog process to find optimal and near-optimal solutions to the energy function (Eq. (\ref{energyfunction})). A quantum annealing algorithm for controlling this process has four components:

\begin{itemize}
\item An {\em initial Hamiltonian} ${\cal H}_{I}$,  which describes initial conditions.  
\item The {\em problem Hamiltonian} ${\cal H}_{P}$ described above. 
\item A pair of {\em path functions}  $A(s)$, $B(s)$ that control the transition from ${\cal H}_I$ to ${\cal H}_P$ over a 
time interval $s: 0 \rightarrow 1$.  In current QPUs these functions are related by $B(s) = 1 - A(s)$.  
\item A parameter $t_{a}$ that specifies total time for the transition in microseconds. 
\end{itemize} 


Quantum annealing uses an adiabatic quantum evolution approach to approximate solutions of the energy function $E(S)$ given by Eq. (\ref{energyfunction}). This is done by traversing from the ground state of an initial Hamiltonian ${\cal H}_{I}$ to a ground state of a final Hamiltonian ${\cal H}_{P}$. According to this scheme, a time dependent Hamiltonian is defined as
\begin{equation}
{\cal H}(t) = A(\tau){\cal H}_{I} + B(\tau){\cal H}_{P}
\label{sec5.eq04}
\end{equation}
where $\tau=t/t_a$ for $0\leq t \leq t_a$ and $t_a$ is the total annealing time. Usually, the ground state of initial Hamiltonian ${\cal H}_{I}$ is easy to prepare and the ground state of the final Hamiltonian ${\cal H}_{P}$ codifies the solution of our problem. In the current generation of the D-Wave 2000Q processor,  functions $A(\tau)$ and $B(\tau)$ are defined so that, at time $\tau=0$, the influence of Hamiltonian ${\cal H}_{I}$ is predominant against ${\cal H}_{P}$. As time evolution goes from $\tau=0$ to $\tau=1$, the influence of Hamiltonian ${\cal H}_{P}$ increases while  ${\cal H}_{I}$ fades away. 



Quantum annealing is the key component of a computation that takes place in stages as follows. 
\begin{enumerate}
\item {\bf Programming/Initialisation.} The weights $(h_i, J_{ij})$ are loaded onto the control system and
qubits are placed in an initial superposition state according to ${\cal H}_I$.   

\item {\bf Anneal.}  The quantum particle process makes a transition from ${\cal H}_I$ to ${\cal H}_P$ according to $A(s), B(s)$, over time $t_{a}$.  

\item {\bf Readout.} At the end of the transition, qubits have spin states according to ${\cal H}_P$ which matches $E(s)$.  Qubit values are read, 
yielding solution $S$ to the input.   

\item {\bf Resampling.}  Since any quantum computation is probabilistic,  there is always a positive (sometimes significant) probability 
that the computation does not finish in ground state.  Given the relatively high initialisation times,  it is cost-effective to repeat the 
anneal-readout cycle many times per input.  
\end{enumerate} 

In current-model D-Wave  systems the initial Hamiltonian is fixed;   the problem Hamiltonian, anneal time $t_{a}$, and the number of resampling steps $R$, are supplied by the user.  Beginning with the D-Wave 2000Q system, the user can also modify the transition by specifying {\em anneal path offsets}, which are deviations from the default anneal path determined by $A(s)$ and $B(s)$.

\begin{figure}
\begin{centering}
\includegraphics[height=1.8in]{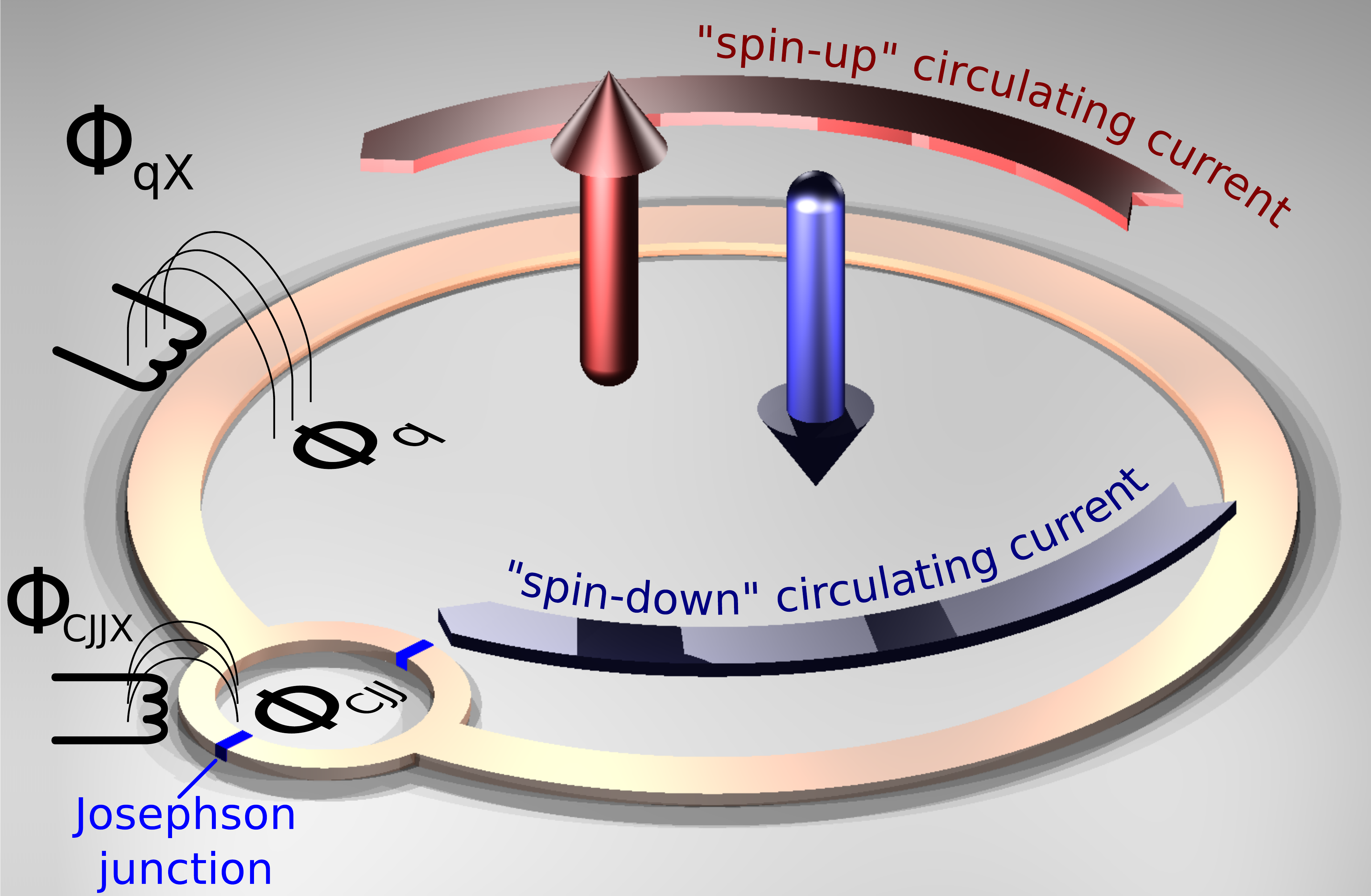}  \\
\end{centering} 
\caption{Each qubit is a loop of niobium controlled by Josephson junctions and biases represented by $\Phi$s in this diagram.  
The direction of current in the loop controls the state of the qubit, which may be spin-up or spin-down.   When cooled to temperatures below 20 mK,  niobium loops can be placed in superposition, with current running in both directions, corresponding to a superposition of spin-up and spin-down states. } 
\label{fluxqubit} 
\end{figure} 

\begin{table}[t]
\caption{Typical computation times}
\label{computationtimes}
\begin{center}
\begin{tabular}{|cccc|}
\hline
 $t_{program}$ & $t_{a}$  & $t_{read}$ & $ T(1000)$ \\ \hline
  9 ms                &$\geq 5 \mu$s    & $120 \mu$s  &  149 ms \\ \hline  
\end{tabular}
\end{center}
\end{table}

The QPU requires total time 
$T(R) = t_{program} + R(t_{a} + t_{read})$
to return a sample of $R$ solutions to one input instance.  Table (\ref{computationtimes}) shows typical component times, and the time to sample $R=1000$ solutions,  for D-Wave 2000Q systems.  Note that computation time does not depend on input size,  and the annealing  step is just a tiny fraction of $T(R)$. 
\\

{\bf Coping with control errors.}  Although SAPI accepts floating point numbers for input weights $(h_i, J_{ij})$, the digital values must be translated to analog control signals, which can introduce precision errors.  Also, although the QPU runs in a chamber that is highly shielded and cryogenically cooled,  some energy gets through.  While thermal energy can be helpful to  the quantum computation \cite{dickson},  it can also be a source of error.  Whether these errors affect the quality of solutions returned by the QPU is highly dependent on the input instance -- or more precisely on the output space specified by the instance.  Very generally speaking,  if the set of optimal solutions $S_0$ has energy    $E(S_0)$ that is well-separated from energies of other solutions, there is no problem;  if energies are similar, the QPU may  have trouble distinguishing optimal from near-optimal solutions.  
   

A number of heuristic strategies are available for the user to apply preprocessing (modifying inputs),  or postprocessing (modifying outputs)  to reduce the impact of control errors in some cases.  Some strategies are provided as SAPI utilities.  Furthermore,  successive models of D-Wave systems have incorporated new technologies that suppress these types of errors.  Their effectiveness is highly input-dependent, but can be substantial:   \cite{denchev}  reports that because of error suppression, a D-Wave 2X processor solved a problem to optimality 10,000 times faster than predictions based on the previous year's D-Wave Two processor. 

\subsection{The technology stack}

\begin{figure}
\begin{centering}
\includegraphics[height=1.9in]{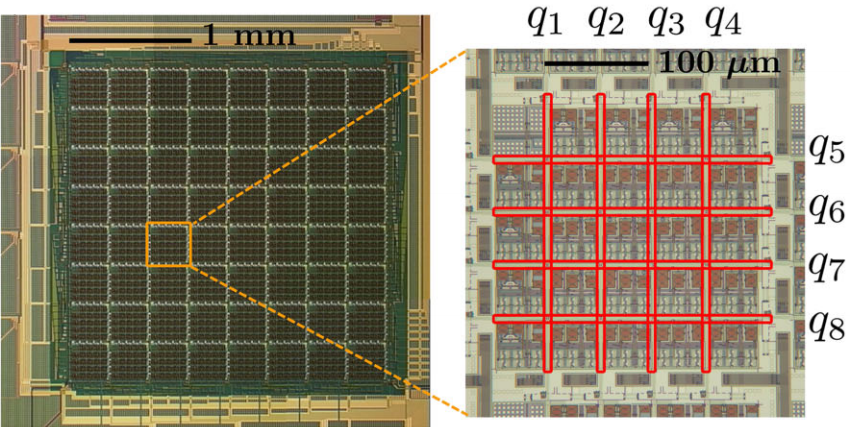}  \\
\end{centering}
 \caption{A photograph of an $8 \times 8$ Chimera graph with a single cell highlighted. The qubits are laid out in long thin loops (red)  connected by couplers (not seen here) at their intersections.  Control devices fill the spaces between qubits.  }  
 \label{chimera-photo} 
 \end{figure}

This section presents an overview of the D-Wave technology stack. We highlight three technologies:  
the superconducting flux qubits that carry quantum state;  the Chimera connection topology;  and the control systems for reading, writing, and manipulating qubit states.   Fig.~(\ref{mounting}) shows  a QPU chip (about the size of a thumbnail), the black diamond at the bottom of its mounting apparatus.  The bundled wires carry control signals and data between the QPU and the control processor. 
\\

\newpage{}
{\bf Qubits and couplers.} A QPU holds an arrangement of {\em superconducting flux qubits}, specifically CCJJ rf-SQUIDS.\footnote{Compound-compound Josephson junction, radio frequency superconducting quantum inference devices.}   Each qubit is made of niobium (Nb) and is in the shape of a double ring interrupted and controlled by two Josephson junctions (Fig.~(\ref{fluxqubit})).  The flux (flow of current) in the ring depends on {\em control biases} represented here by $\Phi$ values.  Current can flow clockwise, counterclockwise, or, when the qubit is in quantum superposition, in both directions at once.  Flux creates magnetic forces represented as spin-up or spin-down, or a combination, which represent qubit superposition states.

 Couplers, which form connections between pairs of qubits, are also made from loops of niobum, but are simpler and have less control  circuitry than qubits.  Qubits and couplers are etched onto silicon wafer circuits containing four Nb-layers separated by insulation. 
 
 Harris et al. \cite{harris} describe some pros and cons of choosing to fabricate with superconducting flux qubits as opposed to other options.  One drawback is that the superconductivity necessary for niobium to experience quantum superposition is only possible when the material is cryogenically frozen to temperatures below 20 mK.  Also, although this technology is robust against many types of noise,  both external and internal  (crosstalk), it is sensitive to electromagnetic interference.  Robustness makes large scale (thousands of qubits) deployment possible,  but on the other hand the computation must be protected from the so-called energy bath of the ambient environment. Thus the QPU can only operate in a shielded chamber where the temperature is below 15 mK, the magnetic field is 50 thousand times less than that Earth's magnetic field, and atmospheric pressure is 10 billion times less than outside. 

One important feature of the rf-SQUIDS deployed in D-Wave QPUs is that they allow a high degree of  tunability, making it  possible to correct for some types of fabrication errors that are inevitable even under state-of-the-art fabrication standards. Tunability also means that qubits are runtime-programmable.  Each qubit has six ``control knobs''  that are used to adjust its physical properties in real time.  This feature produces acceptable levels of consistency and regularity during the quantum annealing process.       
\\

{\bf Connection topology}. Fig.~(\ref{chimera-photo}) shows a photograph of qubits forming a $C_8$ --- an $8 \times 8$ grid of Chimera cells --- with one cell expanded. The qubits (red) form long thin loops arranged in criss-cross fashion,  with couplers (not shown) at every intersection. Control  circuitry is located in the grid interstices of each cell.

 \begin{figure} 
\begin{picture}(900,130)(-35,0)
{
\color{red}
\put(65,25){\oval(120,5)}
\put(65,50){\oval(120,5)}
\put(65,75){\oval(120,5)}
\put(65,100){\oval(120,5)}

\put(25,64){\oval(5,120)}
\put(50,64){\oval(5,120)}
\put(75,64){\oval(5,120)}
\put(100,64){\oval(5,120)}
} 
\newsavebox{\loopy}
\savebox{\loopy}
 (100, 10)[bl]{ %
 \color{blue} 
\multiput(0,0)(0,25){4}{\oval(10,10)[br]} 
\multiput(0,0)(0,25){4}{\oval(15,15)[br]}
\multiput(0,-7.6)(0,25){4}{\line(0,1){3}}
\multiput(4.6,0)(0,25){4}{\line(1,0){3}}

}

\newsavebox{\ellcouple}
\savebox{\ellcouple}(20,20)[bl]{%
\color{blue}
\linethickness{1pt}
\put(0,0){\line(0,1){15}}
\put(5,5){\line(0,1){10}}
\put(0,15){\line(1,0){5}}
\put(0,0){\line(1,0){15}}
\put(5,5){\line(1,0){10}}
\put(15,0){\line(0,1){5}}

}

\multiput(22,22)(25,0){4}{\usebox{\ellcouple}}
\multiput(22,48)(25,0){4}{\usebox{\ellcouple}}
\multiput(22,72)(25,0){4}{\usebox{\ellcouple}}
\multiput(22,97)(25,0){4}{\usebox{\ellcouple}}

\color{blue}
\multiput(126,25)(0,25){4}{\circle{8}}
\multiput(25,125)(25,0){4}{\circle{8}}


\newsavebox{\abigcell}
\savebox{\abigcell}(100,100)[bl]{
{
\color{blue} 
\put(30,0){\line(1,1){45}}    
\put(30,0){\line(-1,1){45}}
\put(30,0){\line(1,3){16}}
\put(30,0){\line(-1,3){16}}

\put(30,30){\line(1,1){16}}   
\put(30,30){\line(-1,1){16}}
\put(30,30){\line(3,1){45}}
\put(30,30){\line(-3,1){45}}

\put(30,90){\line(1,-1){45}}    
\put(30,90){\line(-1,-1){45}} 
\put(30,90){\line(1,-3){16}}
\put(30,90){\line(-1,-3){16}}

\put(30,60){\line(1,-1){16}}   
\put(30,60){\line(-1,-1){16}}
\put(30,60){\line(3,-1){45}}
\put(30,60){\line(-3,-1){45}}
} 

} 
\put(230,10){\usebox{\abigcell}}

\color{blue} 
\multiput(220,55)(30,0){4}{\line(0,1){75}}
\multiput(264,10)(0,30){4}{\line(1,0){70}}

\color{red} 
\multiput(264,10)(0,30){4}{\circle*{8}}  
\multiput(220, 55)(30,0){4}{\circle*{8}}  

\end{picture} 
\caption{One Chimera Cell.  (left) Physical layout of  qubits and couplers. (right) Representation of qubits (red) and couplers (blue) as vertices and edges in a Chimera graph.} 
\label{chimera-cell} 
\end{figure}
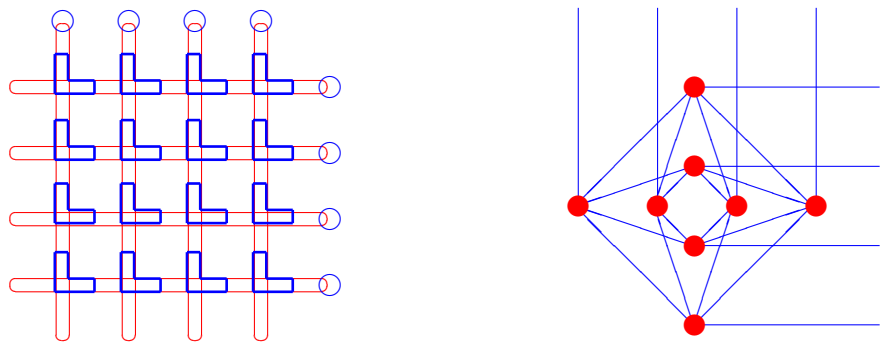 


Fig.~(\ref{chimera-cell}) shows the relationship between the physical layout of qubits (red) and couplers (blue),  and their graphical representation.  Each Chimera cell forms a complete bipartite graph on eight vertices. In addition,  the small blue circles in the layout correspond to the extra blue edges in the graph,  which connect vertices to neighbours in adjacent cells to north and east  (similar connections to south and west are not shown).    

\newpage{}
 
Bunyk et al. \cite{bunyk} list several features that make the Chimera structure attractive to implement, including:  it is robust against crosstalk and noise,  scales well,  and can be fabricated with just two qubit layers.  Ising model problems remain $\mathsf{NP-hard}$ when restricted  to this topology,  because it is non-planar.  Some theorems are available (\cite{choi1,choi2}) to show that this topology supports compact minor-embeddings:  for example a complete graph on $k$ vertices can be minor-embedded into the upper diagonal  of a $C_{k^2}$, which is optimal over a class of physically feasible alternatives.   
 
That being said, Chimera is not the only connectivity structure that could be considered.  Future QPU topologies will exhibit  
higher qubit degree and richer connectivity structure than the current design.    
\\

{\bf Control circuitry.}  As mentioned previously,  each qubit is connected to six dynamic control knobs, which  
are digital-to-analog controllers (DACs) that carry the signals ---called biases --- from the control processor.  For each qubit $q_i$, one DAC carries the bias corresponding to $h_i$,  and the other five are used to tune and correct the qubit state with respect to certain variations from specifications that arise in fabrication.

To load a problem onto the QPU chip, the input weights $(h_i, J_{ij})$ are transferred and latched by DACs.  The control processor uses DACs to initialise qubit  states by ``dialing up'' the scale of ${\cal H}_I$ to the maximum level,  corresponding to path function $A(s)$ at $s=0$,  and \lq \lq dialing down''  the scale of ${\cal H}_P$ to the minimum level, corresponding to $B(s)$ at $s=0$.  The qubits relax into their superposition states according to the inital Hamiltonian,  essentially ignoring the problem Hamiltonian.  The anneal process involves adjusting the scales of ${\cal H}_I$ and ${\cal H}_P$  over time, according to $A(s)$ and $B(s)$.  

At the end of the computation it is necessary to measure qubit flux and store the result as output.  Berkley et al. \cite{berkley}  describe
the qubit-addressible readout circuitry employed in D-Wave QPUs.  Each SQUID qubit is connected to a direct current (dc) SQUID, which has 
a voltage state that depends on the flux imparted by the qubit and which can be read by applying a small voltage pulse.   It does not work to 
simply couple these two SQUIDS together,  because even low-pulse voltages can interefere and alter the flux states of nearby qubits.   The remedy 
described in \cite{berkley} involves inserting a third SQUID called a quantum flux parametron (QFP) that serves as a buffer and allows the dc SQUID to 
safetly latch the qubit state. 


\section{Programming a D-Wave system}
\label{william}

\begin{figure}[ht]
\centering
\includegraphics[height=5cm]{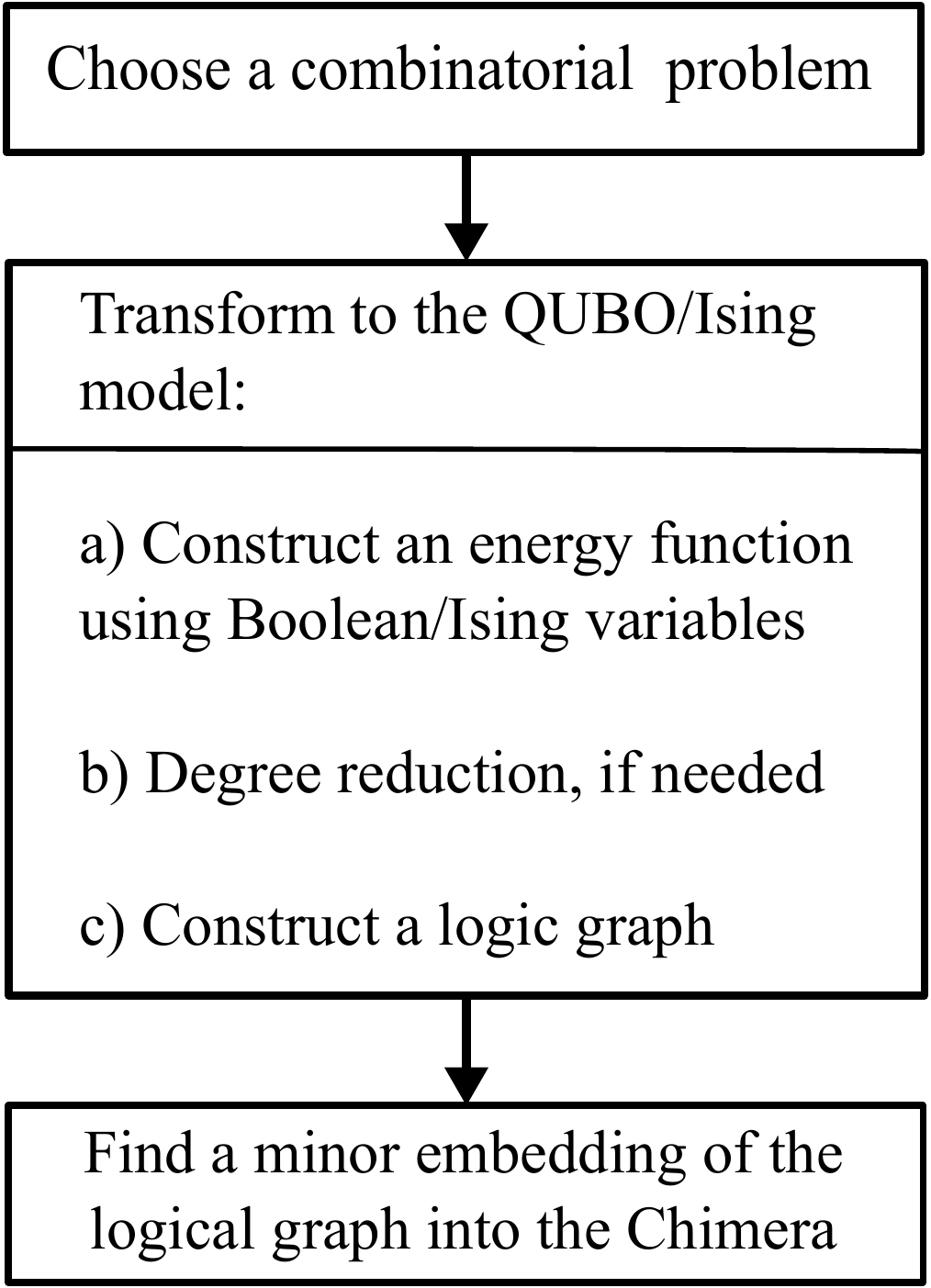}
\caption{Main steps for programming a D-Wave system.}
\label{sec5.fig01}
\end{figure}

The purpose of this section is to present in detail how programs developed to run on a D-Wave system can be created. We start by analysing the mathematical techniques used on the steps shown in the general scheme (Fig.~(\ref{sec5.fig01})) to be followed in order to develop quantum annealing-based algorithms to run on a D-Wave system. Then, we present two $\mathsf{NP-hard}$ problems and corresponding quantum annealing-based algorithms to be executed on a D-Wave quantum Processing Unit (QPU). The $\mathsf{NP-hard}$ problems we have chosen are from mathematical logic and graph theory: the maximum satisfiability and the minimum multicut problems, respectively. Results for the maximum satisfiability problem were produced using the D-Wave SAPI simulation software while results for the minimum multicut problem were produced by running quantum annealing-based algorithms on a D-Wave 2X system.

\subsection{The QUBO and Ising models}

A quadratic unconstrained pseudo-Boolean optimisation (QUBO) problem is a function of the form
\begin{equation}
f_{ue}(X) = \sum_j u_j x_j + \sum_{i<j} e_{ij}x_i x_j
\label{sec5.eq01}
\end{equation}
where $X=(x_1,\dots,x_n)$ is a vector of binary-valued variables, $u\in\mathbb{R}^n$ and $e\in\mathbb{R}^{n\times n}$ are a vector and a matrix of real coefficients, respectively. Since finding the minimisation of a QUBO problem is an $\mathsf{NP-hard}$ problem,  the best algorithm known must evaluate $2^n$ possible assignments to the variables in X to find the minimum cost of $f_{ue}$.  A QUBO model is related to the Ising model  via variable substitution $x_i = (1+s_i)/2$ for $i=1,\dots,n$. 

\subsection{Degree reduction}



The formulation of $\mathsf{NP-hard}$ problems using Boolean variables tends to produce terms with more than two interacting variables. Those  representations are known as pseudo-Boolean functions and are defined as follows

\begin{equation}
f(X) = \sum_{S\subseteq \{1,\dots,n\}} \alpha_S \prod_{j\in S} x_j
\label{sec5.eq05}
\end{equation}
where $\alpha_S$ are real coefficients. The degree of function $f$, $\mbox{deg}(f)$, is equal to the cardinality of the largest subset $S\subseteq \{1,\dots,n\}$ for which $\alpha_S\neq 0$ while the size of $f$ is the total number of variable occurrences (note that QUBO coincides with functions $f$ with $\mbox{deg}(f)\leq 2$). Every pseudo-Boolean function $f$ expressed as in Eq. (\ref{sec5.eq05}) can be reduced in polynomial time~\cite{Boros2002}, with respect to the size of $f$, to a QUBO function $f_{ue}$ in $m$ variables, with size polynomially bounded by $\mbox{size}(f)$, so that $\min_{y\in \{0,1\}^m} f_{ue}(y) = \min_{x\in \{0,1\}^n} f(x)$. 


\newpage{}

Among existing reduction methods \cite{Boros2002,Boros2014,Anthony20161}, we favour those that produce as few new variables as possible (reducing a pseudo-Boolean function to a quadratic one using the smallest number of variables is $\mathsf{NP-hard}$ \cite{Boros2002}). We introduce two well-known reduction methods, the Freedman~\cite{Freedman05} and Ishikawa~\cite{Ishikawa11} methods:




\begin{enumerate}
\item {\em Freedman method}. It is applied to negative terms and requires only one new variable. For any arbitrary negative monomial of degree $d$, Eq. (\ref{sec5.eq06}) holds:
\begin{equation}
-\prod_{j=1}^d x_j = \min_{w\in\{0,1\}} w\left((d-1) - \sum_{j=1}^d x_j\right)
\label{sec5.eq06}
\end{equation}
where $w$ is a new Boolean variable.

\begin{example}
Let $f=x_1x_2 - x_1x_2x_3$ be a cubic pseudo-Boolean function. Then, $f$ can be written as $f=x_1x_2 + \min_{x_4\in\{0,1\}} x_4(2-x_1-x_2-x_3)$ where $x_4$ is a new variable. Omitting the min function, one can obtain a new quadratic function $f'=x_1x_2 + x_4(2-x_1-x_2-x_3)$ such that $\min_{x\in\{0,1\}^3} f(x)=\min_{y\in\{0,1\}^4} f'(y)$. 
\end{example}

\item {\em Ishikawa method}. This method requires $\lfloor \frac{d-1}{2}\rfloor$ new variables to reduce a positive monomial of degree $d$. Formally, let $t(x) = x_1x_2\cdots x_d$ of degree $d$, $k=\lfloor\frac{d-1}{2}\rfloor$ and $w = (w_1,w_2,\dots, w_k)$ be a vector of $k$ new variables. Furthermore, we define
\begin{eqnarray}
S_1=\sum_{j=1}^{d}x_j, \ \ S_2=\sum_{1\leq i< j\leq d}x_ix_j, \ \  W_1=\sum_{j=1}^{k}w_j, \ \text{and}\  W_2=\sum_{j=1}^{k}(4j-1)w_j. \nonumber
\end{eqnarray}

Therefore, it is satisfied that: 
\begin{equation}
t(x) =
  \begin{cases}
    S_2 +  \min\limits_{w \in \{0,1\}^k} \{W_2 - 2W_1S_1\}       & \quad \text{if } d \text{ is even,} \\
    S_2 +  \min\limits_{w \in \{0,1\}^k}  \{W_2 - 2W_1S_1 +w_k(S_1-d+1)\}  & \quad \text{if } d \text{ is odd}. \\
  \end{cases}
  \label{sec5.eq07}
\end{equation}

\begin{example}
Let $f=x_1x_2 + x_1x_2x_3$ be a cubic pseudo-Boolean function. Then, $f$ can be written as $f = 2x_1x_2 + x_1x_3 + x_2x_3 + \min_{x_4\in\{0,1\}} x_4 (1-x_1-x_2-x_3)$ where $x_4$ is a new variable. Omitting the min function, one can obtain a new quadratic function $f' = 2x_1x_2 + x_1x_3 + x_2x_3 + x_4 (1-x_1-x_2-x_3)$ such that $\min_{x\in\{0,1\}^3} f(x)=\min_{y\in\{0,1\}^4} f'(y)$. 
\end{example}
\end{enumerate}

\subsection{Minor embedding}

As described in section \ref{dwavecathy}, the Chimera graph $\mathcal{G}_{M,N,L}$ consists of a $M\times N$ two-dimensional lattice of blocks. Each block is composed of $2L$ qubits for a total of $2MNL$ qubits. The D-Wave system was designed to solve instances of Ising problems that can be mapped to the Chimera topology. For any Ising energy function $E(S)$ on $n$ variables as given in Eq. (\ref{energyfunction}),  we associate a logical weighted graph $G_{\mbox{\tiny Ising}}=(V_{\mbox{\tiny Ising}},E_{\mbox{\tiny Ising}})$  where $V_{\mbox{\tiny Ising}}=\{1,\dots,n\}$ and $E_{\mbox{\tiny Ising}}=\{\{i,j\} \mid J_{ij}\neq 0 \}$ such that for each $j\in V_{\mbox{\tiny Ising}}$ and $\{i,j\}\in E_{\mbox{\tiny Ising}}$,  weights $h_j$ and $J_{ij}$ are respectively assigned. Thus, in order to find the ground state of function $E(S)$ using a D-Wave system, we must map the logical graph $G_{\mbox{\tiny Ising}}$ into the Chimera.

\newpage{}

Usually, it is not possible to find a one-to-one mapping between logical variables and physical qubits into the Chimera, i.e.  $G_{\mbox{\tiny Ising}}$ is not always a subgraph of $\mathcal{G}_{M,N,L}$. A strategy to find an equivalent subgraph into the Chimera of a given logical graph $G_{\mbox{\tiny Ising}}$ was proposed by Kaminsky et al. \cite{Kaminsky2004a,Kaminsky2004b} in which each vertex in $G_ {\mbox{\tiny Ising}}$ is mapped to a connected subtree of qubits that represent a logical variable. This problem is called the minor embedding and it has been introduced in section \ref{dwavecathy}.


\begin{prob}
Let $\mathcal{G}_{M,N,L}$ be a Chimera graph of dimension $(M,N,L)$ and $G_{\mbox{\tiny Ising}}=(V_{\mbox{\tiny Ising}},E_{\mbox{\tiny Ising}})$ a logical graph. Find a subgraph in $\mathcal{G}_{M,N,L}$ such that
\begin{enumerate}
\item Each vertex $j\in V_{\mbox{\tiny Ising}}$ is mapped to a connected subtree $T_j$ in $\mathcal{G}_{M,N,L}$.
\item Each edge $\{i,j\}\in E_{\mbox{\tiny Ising}}$ must be mapped to at least one coupler in $\mathcal{G}_{M,N,L}$.
\end{enumerate}
\end{prob}

Table (\ref{tabla01}) shows examples of minor embedding of the complete graphs (in which there exists an edge between every pair of vertices) $K_3$, $K_4$ and $K_5$. The embedding for the graph $K_3$ can be explained as follows: vertices 1, 2 and 3 in $K_3$ are mapped to the qubits $q_1$, $q_5$ and $q_6$. Using these qubits, the edges $\{1,2\}$ and $\{1,3\}$ are mapped to the couplers $\{q_1,q_2\}$ and $\{q_1,q_3\}$, respectively. To map the edge $\{2,3\}$, we use an extra qubit $q_2$ to create a chain between $q_2$ and $q_6$, representing vertex 3 and the coupler $\{q_2,q_5\}$ is added. Qubits $q_2$ and $q_6$ are joined with a coupler to represent the same vertex 3. The minor embedding of the graphs $K_4$ and $K_5$ can be explained using a similar reasoning. Notice that we require only one block of the Chimera to find a minor embedding of the graphs $K_3$, $K_4$ and $K_5$.

\begin{table}[t]
\caption{Examples of minor embedding into the $(1,1,4)$-Chimera of the complete graphs $K_3$, $K_4$ and $K_5$. The numbers and colors of the vertices in the graph are the same as in the minor embedding. Bold black lines correspond to the mapped edges and bold color lines correspond to chain of qubits.}
\begin{center}
\begin{tabular}{| >{\centering\arraybackslash}m{0.65in} | >{\centering\arraybackslash}m{1.4in} | >{\centering\arraybackslash}m{1.32in} | >{\centering\arraybackslash}m{1.32in} |}
\hline
\multirow{2}{*}{Graph}
 &   \dummyfigure{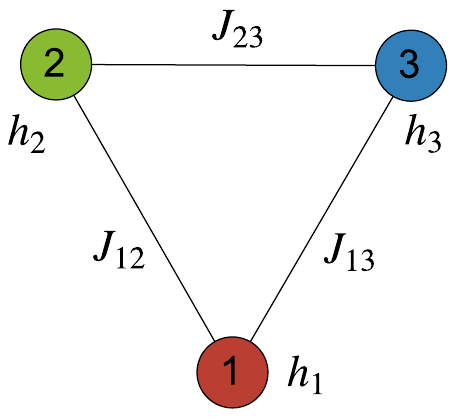}{2.4}                 &  \dummyfigure{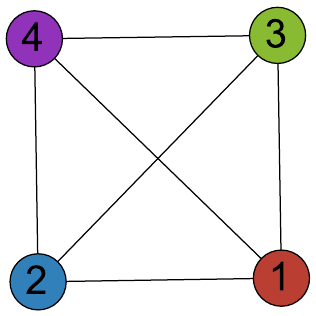}{2.25}      &     \dummyfigure{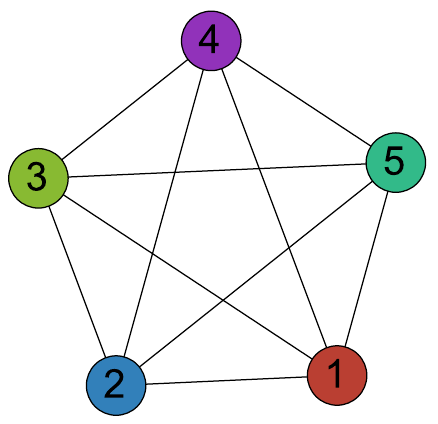}{2.7}      \\
       & $K_3$ & $K_4$ & $K_5$ \\
\hline
Minor embedding & \dummyfigure{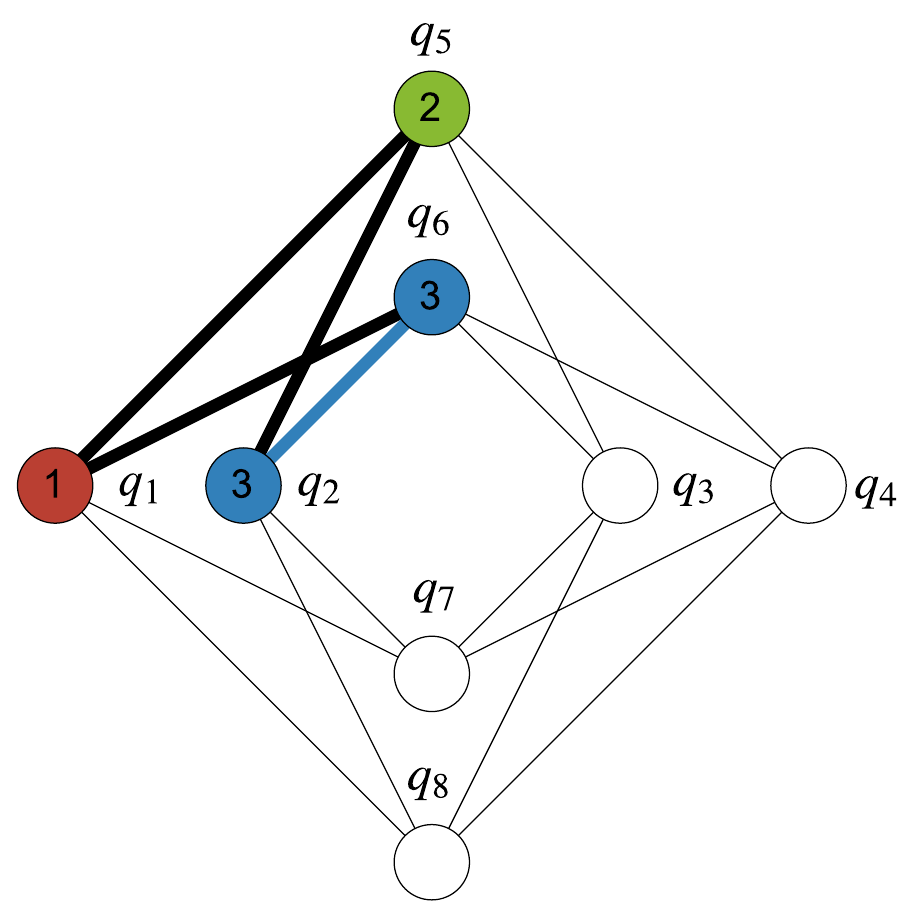}{3.8} & \dummyfigure{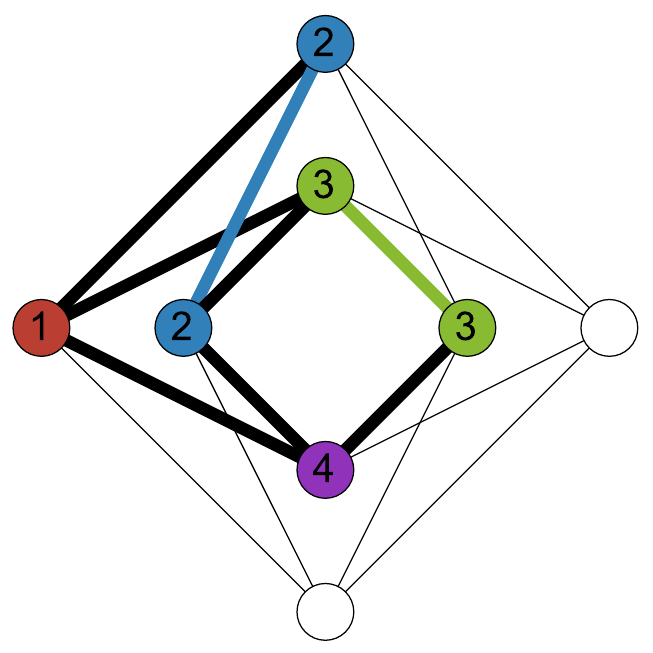}{3.5} & \dummyfigure{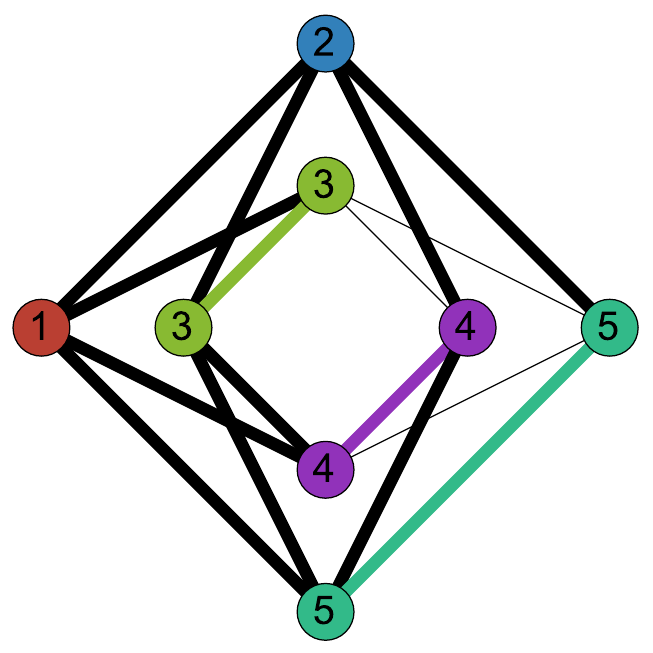}{3.5}  \\
 \hline
\end{tabular}
\end{center}
\label{tabla01}
\end{table}


After a minor embedding is found for a given graph $G_{\mbox{\tiny Ising}}$, we need to assign the weights to the qubits and coupler without changing the representation of the problem. The following procedure accomplishes this requirement: (1) for each $i\in V_{\mbox{\tiny Ising}}$, let $T_i=(V_i,E_i)$ be the connected tree that represents vertex $i$; (2) add a weight $-M$ for each connection in $E_i$ where $M>0$; (3) add a weight $h_i^j$ for each qubit $q_i^j\in V_i$ such that $h_i^1+\cdots +h_i^k=h_i$ with $k=|V_i|$; (4) add the weight $J_{ij}$ to the coupler $\{q_{i_1},q_{i_2}\}$ used to represents edge $\{i,j\}\in E_{\mbox{\tiny Ising}}$.

To exemplify the previous procedure, let us consider the complete graph $K_3$ in Table~(\ref{tabla01}). The graph $K_3$ represents the following Ising problem

\begin{equation}
E(S) = h_1s_1 + h_2s_2 + h_3s_3 + J_{12}s_1s_2 + J_{13}s_1s_3 + J_{23}s_2s_3,\nonumber
\end{equation}
and its equivalent problem, according to its minor embedding, is
\begin{equation}
{E'}(q) = h_1q_1 + h_2q_5 + (h_3^1q_2 + h_3^2q_6) + J_{12}q_1q_5 + J_{13}q_1q_6 + J_{23}q_2q_5 - Mq_2q_6 \nonumber
\end{equation}
where $h_3^1 + h_3^2=h_3$ and $M>0$.

The value of $M$,  called the {\it chain strength},  should be large enough to ensure that the ground state of $E'(q)$ corresponds to $\text{spin}_2 = \text{spin}_6$.   Choosing $M$ too large may reduce performance of the quantum processor by introducing control errors as described in the previous section.

\subsection{Application to the max-SAT problem}
 
 As described in section \ref{computerscience}, the maximum satisfiability (max-SAT) is an optimisation version of SAT that consists of determining the maximum number of satisfied clauses in a given SAT formula. Let us describe a mathematical formulation of the max-SAT problem in terms of a QUBO function. Given a formula  $\Phi=\bigwedge_{j=1}^m C_j$ over $n$ Boolean variables, let $h_{\Phi}:\{0,1\}^n\rightarrow\mathbb{Z}$ be a function defined as 
\begin{equation}
h_{\Phi} =  h_{C_1}+\cdots+h_{C_m}
\label{sec5.eq08}
\end{equation}
where $h_{C_j}(x)=0$ if $x$ satisfies clause $C_j$, and $h_{C_j}(x)=1$ otherwise, for all assignment $x\in\{0,1\}^n$. $h_{\Phi}$ is used to count the number of unsatisfied clauses. For convenience, and given that any Boolean formula in general   form can be expressed as a max3-SAT problem \cite{arora09}, we assume that clauses are of the form  $C_j=x_{i1}^{\delta_{i1}}\vee x_{i2}^{\delta_{i2}}\vee x_{i3}^{\delta_{i3}}$ (i.e., 3-SAT clauses), where a literal is denoted as $x^{\delta}$ such that $x^0 = \overline{x}$ and $x^1 = x$. Thus, $h_{C_j}$ can be modeled as a cubic pseudo-Boolean function $h_{C_j} = \prod_{k=1}^3 (1 - x_{ik}^{\delta_{ik}})$. By doing several algebraic operations, we have that 
\begin{eqnarray}
h_{C_j} &=&  \alpha_1 +  \alpha_2 x_{i1} + \alpha_3 x_{i2} +  \alpha_4 x_{i3} + \alpha_5 x_{i1} x_{i2} + \nonumber\\ 
&& \alpha_6 x_{i1} x_{i3} +  \alpha_7 x_{i2} x_{i3} +  \alpha_8 x_{i1} x_{i2} x_{i3}
 \label{sec5.eq09}
\end{eqnarray}
whose coefficients $\alpha_1,\dots,\alpha_8$ are determined by the values of $\delta_{i1}$, $\delta_{i2}$ and $\delta_{i3}$. Depending of the chosen literals in each clause and after a simplification of the terms in $h_{\Phi}$, there will be at most $\binom{n}{3}$ cubic terms. 

Now by adding in an extra variable $w$, for each cubic term $\alpha x_{i1} x_{i2} x_{i3}$ in $h_{\Phi}$, we satisfy the following inequality


\begin{equation}
\alpha x_{i1}x_{i2}x_{i3}  \leq
  \begin{cases}
     \alpha[x_{i1}x_{i2} + x_{i1}x_{i3} + x_{i2}x_{i3} +      \\
    w (1 - x_{i1} - x_{i2} - x_{i3} )]  & \quad \text{if } \alpha \text{ is positive},\\
    |\alpha| w ( 2 - x_{i1} - x_{i2} - x_{i3})      & \quad \text{if } \alpha \text{ is negative}.
  \end{cases} 
   \label{sec5.eq10}
\end{equation}


This relation  was obtained using the Ishikawa and Freedman reduction  methods. Hence by substituting the representation of $\alpha x_{i1} x_{i2} x_{i3}$ given in Eq. (\ref{sec5.eq10}), for each cubic term in $h_{\Phi}$ into Eq. (\ref{sec5.eq09}), we conclude that the minimisation of $h_{\Phi}$ over all assignments $(x_1,\dots,x_n,w_1,\dots,w_{l})\in\{0,1\}^{n+l}$ where $1\leq l \leq\binom{n}{3}$, is equivalent to the max-SAT problem.

\newpage{}

As an example, consider the following formula $\Phi$ with a unique satisfying assignment, namely $(0, 0, 1, 1, 1, 1, 1, 1, 1)$,  over $n=9$ variables and $m=39$ clauses:
\begin{eqnarray}
\Phi &=&
(\overline{x}_3 \vee x_6 \vee x_9 ) \wedge
(x_2 \vee \overline{x}_4 \vee x_6 ) \wedge
(\overline{x}_1 \vee \overline{x}_7 \vee \overline{x}_8 )\wedge
(\overline{x}_3 \vee x_4 \vee x_9 ) \wedge \nonumber\\
&& (\overline{x}_1 \vee \overline{x}_5 \vee \overline{x}_8 )\wedge
(x_4 \vee \overline{x}_5 \vee \overline{x}_8 )\wedge
(x_2 \vee x_3 \vee x_5 ) \wedge
(\overline{x}_4 \vee \overline{x}_5 \vee x_7 ) \wedge \nonumber\\
&& (\overline{x}_2 \vee x_5 \vee x_6 ) \wedge
(x_1 \vee x_3 \vee \overline{x}_4 )\wedge
(\overline{x}_1 \vee x_5 \vee \overline{x}_9 )\wedge
(x_1 \vee x_3 \vee x_4 ) \wedge \nonumber\\
&& (x_3 \vee \overline{x}_6 \vee \overline{x}_7 )\wedge
(x_4 \vee x_6 \vee \overline{x}_7 )\wedge
(\overline{x}_4 \vee \overline{x}_7 \vee x_8 ) \wedge
(\overline{x}_1 \vee x_2 \vee \overline{x}_4 )\wedge \nonumber\\
&& (x_1 \vee \overline{x}_4 \vee x_7 ) \wedge
(x_6 \vee x_8 \vee x_9 ) \wedge
(x_4 \vee \overline{x}_5 \vee \overline{x}_9 )\wedge
(x_2 \vee \overline{x}_4 \vee x_8 ) \wedge \nonumber\\
&& (\overline{x}_2 \vee \overline{x}_3 \vee \overline{x}_8 )\wedge
(x_1 \vee \overline{x}_5 \vee x_7 ) \wedge
(\overline{x}_2 \vee \overline{x}_3 \vee \overline{x}_5 )\wedge
(\overline{x}_4 \vee x_5 \vee x_7 ) \wedge \nonumber\\
&& (\overline{x}_2 \vee \overline{x}_6 \vee \overline{x}_8 )\wedge
(\overline{x}_4 \vee \overline{x}_6 \vee x_9 ) \wedge
(x_7 \vee \overline{x}_8 \vee \overline{x}_9 )\wedge
(x_2 \vee \overline{x}_3 \vee x_5 ) \wedge \nonumber\\
&& (\overline{x}_2 \vee \overline{x}_4 \vee x_5 ) \wedge
(\overline{x}_1 \vee \overline{x}_4 \vee \overline{x}_6 )\wedge
(x_1 \vee x_4 \vee x_5 ) \wedge
(x_5 \vee \overline{x}_7 \vee \overline{x}_9 )\wedge \nonumber\\
&& (x_1 \vee x_2 \vee x_4 ) \wedge
(\overline{x}_2 \vee x_3 \vee \overline{x}_6 )\wedge
(\overline{x}_4 \vee x_7 \vee \overline{x}_8 )\wedge
(x_1 \vee \overline{x}_3 \vee x_6 ) \wedge \nonumber\\
&& (\overline{x}_3 \vee \overline{x}_4 \vee x_5 ) \wedge
(\overline{x}_1 \vee \overline{x}_3 \vee x_7 ) \wedge
(\overline{x}_1 \vee x_3 \vee x_9 ) 
\label{sec5.eq11}
\end{eqnarray}

Applying  Eq. (\ref{sec5.eq08}) on Eq. (\ref{sec5.eq11}), we have that

\begin{eqnarray}
h_{\Phi} &=& 5 + 2 x_{3} + 2 x_{4} - x_{2} - x_{5} - 2 x_{1} + x_{7} - x_{6} - x_{8} - x_{9} - 2 x_{3}x_{6} - \nonumber\\
&& 2 x_{3}x_{9} +  x_{5} x_{8}  + x_{4} x_{5}  + 2 x_{1}x_{4} - 2 x_{4}x_{7} + x_{6} x_{8}  + x_{6} x_{9}  + 2 x_{8}x_{9} + \nonumber\\
&& x_{5} x_{9}  - x_{5} x_{7}  + x_{7} x_{9}  + x_{1} x_{2}  + x_{3} x_{6} x_{9} + x_{2} x_{4} x_{6} + x_{1} x_{7} x_{8} + x_{3} x_{4} x_{9} \nonumber\\
&& + x_{1} x_{5} x_{8} - x_{4} x_{5} x_{8} + x_{2} x_{3} x_{5} + x_{2} x_{5} x_{6} - x_{1} x_{5} x_{9} - x_{3} x_{6} x_{7} + \nonumber\\
&& x_{4} x_{6} x_{7} - 2 x_{4} x_{7} x_{8} - 2 x_{1} x_{2} x_{4} + x_{1} x_{4} x_{7} - x_{6} x_{8} x_{9} - x_{4} x_{5} x_{9} + \nonumber\\
&& x_{2} x_{4} x_{8} + x_{2} x_{3} x_{8} + x_{1} x_{5} x_{7} + x_{2} x_{6} x_{8} - x_{4} x_{6} x_{9} - x_{7} x_{8} x_{9} - x_{2} x_{4} x_{5} + \nonumber\\
&& x_{1} x_{4} x_{6} - x_{1} x_{4} x_{5} - x_{5} x_{7} x_{9} - x_{2} x_{3} x_{6} + x_{1} x_{3} x_{6} - x_{3} x_{4} x_{5} - x_{1} x_{3} x_{7} + \nonumber\\
&& x_{1} x_{3} x_{9},
\label{sec5.eq12}
\end{eqnarray}
and after the reduction described in Eq. (\ref{sec5.eq10}), $h_{\Phi}$ becomes
\begin{eqnarray}
h_{\Phi}^{\mbox{\tiny qubo}} &=& 5 + 2 x_{3} + 2 x_{4} - x_{2} - x_{5} - 2 x_{1} + x_{7} - x_{6} - x_{8} - x_{9} + x_{10} + x_{11} + \nonumber\\
&& x_{12} + x_{13} + x_{14} + 2 x_{15} + x_{16} + x_{17} + 2 x_{18} + 2 x_{19} + x_{20} + 4 x_{21} + \nonumber\\
&& 4 x_{22} + x_{23} + 2 x_{24} + 2 x_{25} + x_{26} + x_{27} + x_{28} + x_{29} + 2 x_{30} + 2 x_{31} + \nonumber\\
&& 2 x_{32} + x_{33} + 2 x_{34} + 2 x_{35} + 2 x_{36} + x_{37} + 2 x_{38} + 2 x_{39} + x_{40} + x_{3} x_{9}  + \nonumber\\
&& 2 x_{5}x_{8} + x_{4} x_{5}  + 4 x_{1}x_{4} + 2 x_{6}x_{8} + 2 x_{6}x_{9} + 2 x_{8}x_{9} + x_{5} x_{9}  + x_{7} x_{9}  + \nonumber\\
&& x_{1} x_{2}  - x_{3} x_{10}  - x_{6} x_{10}  - x_{9} x_{10}  + 2 x_{2}x_{4} + 3 x_{2}x_{6} + 3 x_{4}x_{6} - x_{2} x_{11}  - \nonumber\\
&& x_{4} x_{11}  - x_{6} x_{11}  + 3 x_{1}x_{7} + 2 x_{1}x_{8} + x_{7} x_{8}  - x_{1} x_{12}  - x_{7} x_{12}  - x_{8} x_{12}  + \nonumber\\
&& x_{3} x_{4}  + x_{4} x_{9}  - x_{3} x_{13}  - x_{4} x_{13}  - x_{9} x_{13}  + 2 x_{1}x_{5} - x_{1} x_{14}  - x_{5} x_{14}  - \nonumber\\
&& x_{8} x_{14}  - x_{4} x_{15}  - x_{5} x_{15}  - x_{8} x_{15}  + 2 x_{2}x_{3} + 2 x_{2}x_{5} + x_{3} x_{5}  - x_{2} x_{16}  - \nonumber\\
&& x_{3} x_{16}  - x_{5} x_{16}  + x_{5} x_{6}  - x_{2} x_{17}  - x_{5} x_{17}  - x_{6} x_{17}  - x_{1} x_{18}  - x_{5} x_{18}  - \nonumber\\
&& x_{9} x_{18}  - x_{3} x_{19}  - x_{6} x_{19}  - x_{7} x_{19}  + x_{6} x_{7}  - x_{4} x_{20}  - x_{6} x_{20}  - x_{7} x_{20}  - \nonumber\\
&& 2 x_{4}x_{21} - 2 x_{7}x_{21} - 2 x_{8}x_{21} - 2 x_{1}x_{22} - 2 x_{2}x_{22} - 2 x_{4}x_{22} - x_{1} x_{23}  - \nonumber\\
&& x_{4} x_{23}  - x_{7} x_{23}  - x_{6} x_{24}  - x_{8} x_{24}  - x_{9} x_{24}  - x_{4} x_{25}  - x_{5} x_{25}  - x_{9} x_{25}  + \nonumber\\
&& 3 x_{2}x_{8} + x_{4} x_{8}  - x_{2} x_{26}  - x_{4} x_{26}  - x_{8} x_{26}  + x_{3} x_{8}  - x_{2} x_{27}  - x_{3} x_{27}  - \nonumber\\
&& x_{8} x_{27}  - x_{1} x_{28}  - x_{5} x_{28}  - x_{7} x_{28}  - x_{2} x_{29}  - x_{6} x_{29}  - x_{8} x_{29}  - x_{4} x_{30}  - \nonumber\\
&& x_{6} x_{30}  - x_{9} x_{30}  - x_{7} x_{31}  - x_{8} x_{31}  - x_{9} x_{31}  - x_{2} x_{32}  - x_{4} x_{32}  - x_{5} x_{32}  + \nonumber\\
&& 2 x_{1}x_{6} - x_{1} x_{33}  - x_{4} x_{33}  - x_{6} x_{33}  - x_{1} x_{34}  - x_{4} x_{34}  - x_{5} x_{34}  - x_{5} x_{35}  - \nonumber\\
&& x_{7} x_{35}  - x_{9} x_{35}  - x_{2} x_{36}  - x_{3} x_{36}  - x_{6} x_{36}  + 2 x_{1}x_{3} - x_{1} x_{37}  - x_{3} x_{37}  - \nonumber\\
&& x_{6} x_{37}  - x_{3} x_{38}  - x_{4} x_{38}  - x_{5} x_{38}  - x_{1} x_{39}  - x_{3} x_{39}  - x_{7} x_{39}  + x_{1} x_{9}  - \nonumber\\
&& x_{1} x_{40}  - x_{3} x_{40}  - x_{9} x_{40}  .  
\label{sec5.eq13}
 \end{eqnarray}


Notice that solving the max-SAT instance $\Phi$ given in Eq. (\ref{sec5.eq11}) is equivalent to optimizing the function $h_{\Phi}^{\mbox{\tiny qubo}}$. To solve the instance using a D-Wave system, convert $h_{\Phi}^{\mbox{\tiny qubo}}$ to an equivalent Ising function through a change of variables $x_i =  (1+s_i)/2$. This yields an instance of the form $s^T J s + h^T s + \text{\em ising\_offset} = x^T Q x$ where $h$ is a column vector containing the equivalent linear Ising coefficients, $J$ is a matrix of equivalent Ising coupling coefficients, and $s$ is the column vector of Ising variables; $x$ is a column vector of Boolean variables, $Q$ is the matrix of QUBO coefficients and {\em ising\_offset} is a constant shifting all Ising energies.

 \begin{figure}[t]
\centering
\includegraphics[height=5cm]{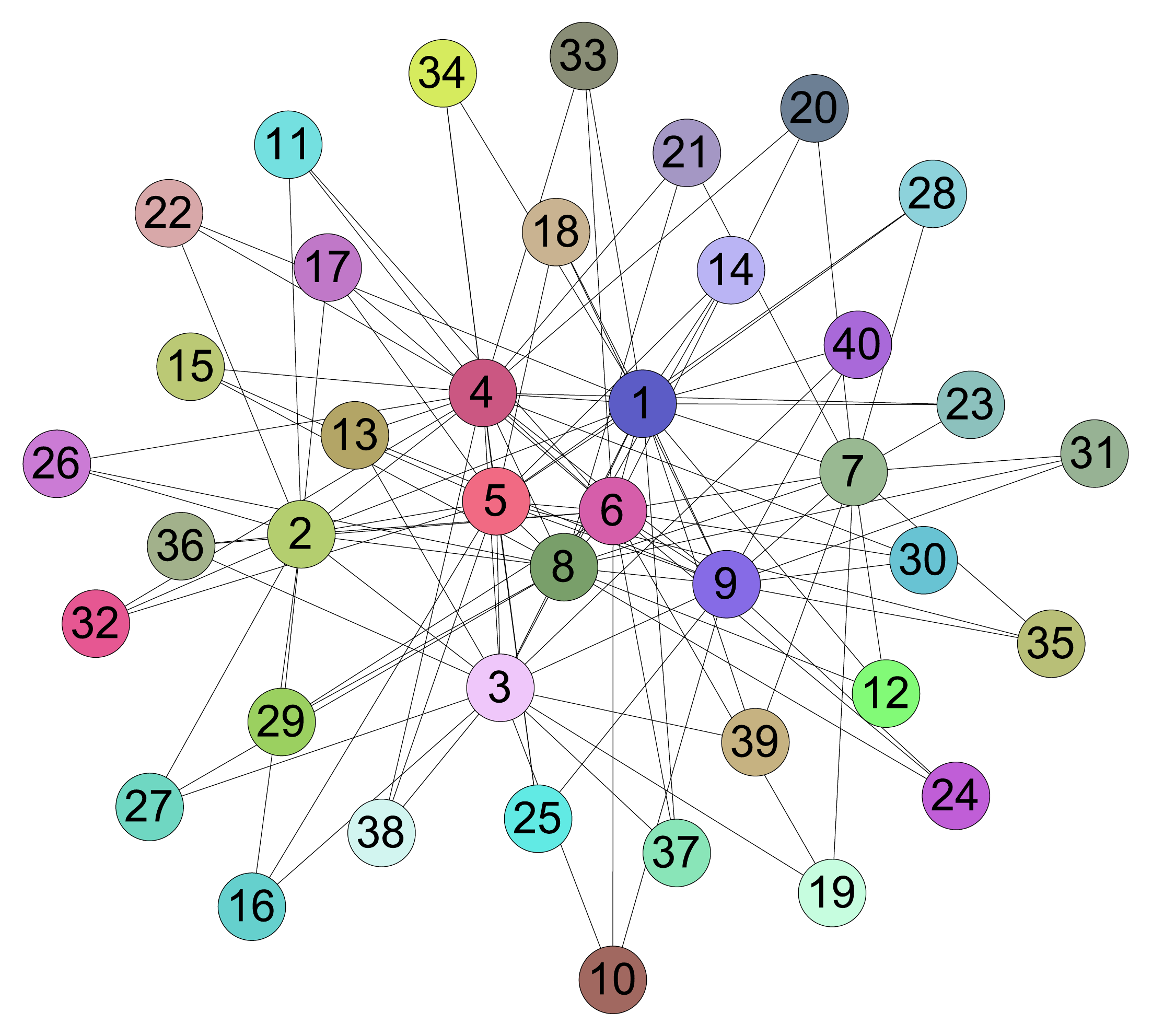}
\includegraphics[height=7cm]{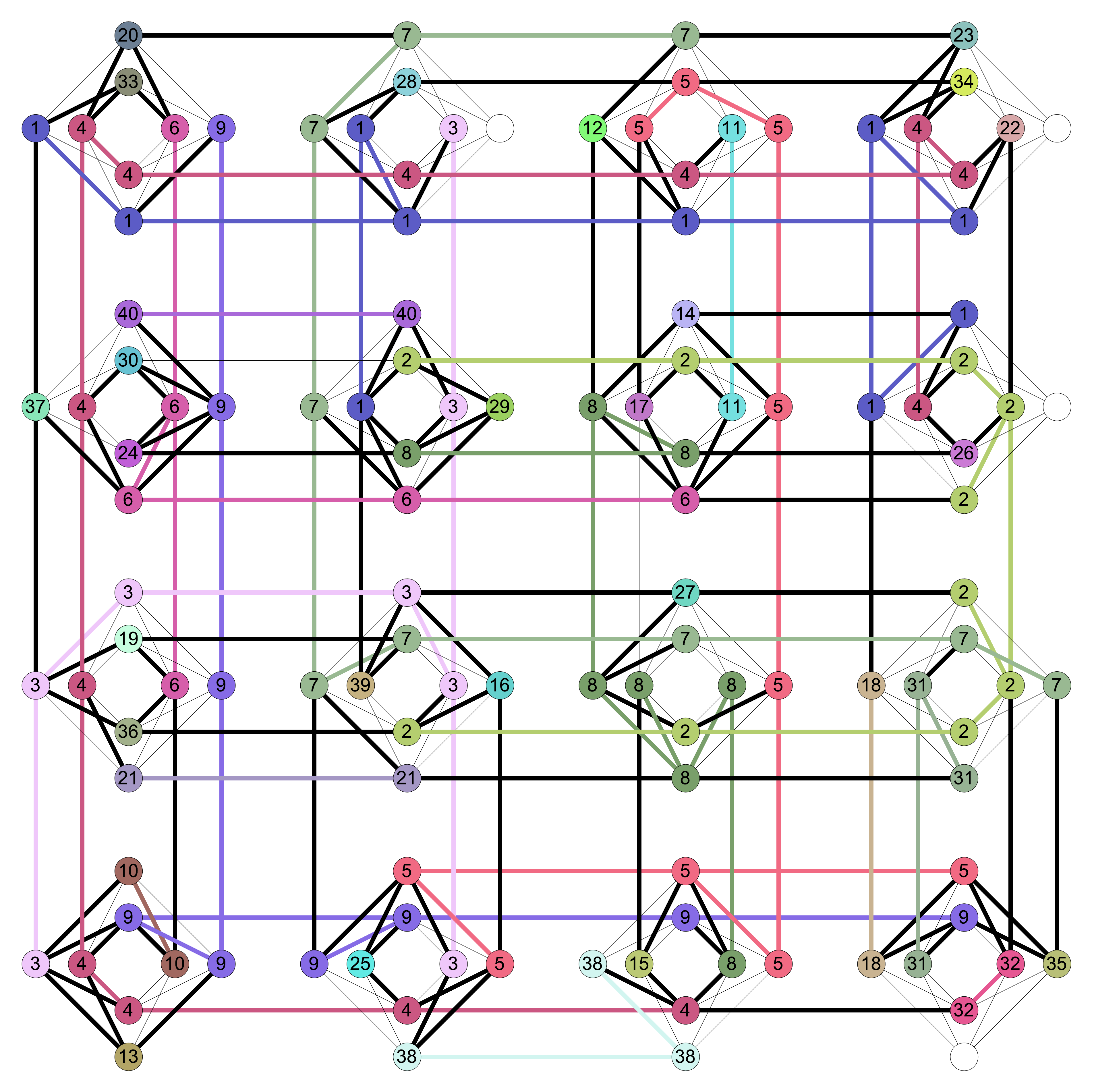}
\caption{(left) logical graph $G_{\mbox{\tiny Ising}}$ corresponding to the QUBO function $h_{\Phi}^{\mbox{\tiny qubo}}$ given in Eq. (\ref{sec5.eq13}) and (right) minor embedding of $G_{\mbox{\tiny Ising}}$ into the Chimera. The logical graph has 40 variables and its minor embedding requires 124 physical qubits. The numbers and colors of the vertices in the logical graph are the same as in the minor embedding. Bold black lines correspond to the mapped edges and bold color lines correspond to chain of qubits.}
\label{sec5.fig03}
\end{figure}

Fig.~(\ref{sec5.fig03})(left) shows the logical graph $G_{\mbox{\tiny Ising}}$ of the equivalent Ising function of   $h_{\Phi}^{\mbox{\tiny qubo}}$ given in Eq. (\ref{sec5.eq13}). The constant {\em ising\_offset} may be ignored since the objective is to minimise both functions. Fig.~(\ref{sec5.fig03})(right) shows a minor embedding of $G_{\mbox{\tiny Ising}}$ into the Chimera. This embedding was obtained using a heuristic algorithm available from the D-Wave SAPI.

\begin{figure}[t]
\centering
\includegraphics[height=5.8cm]{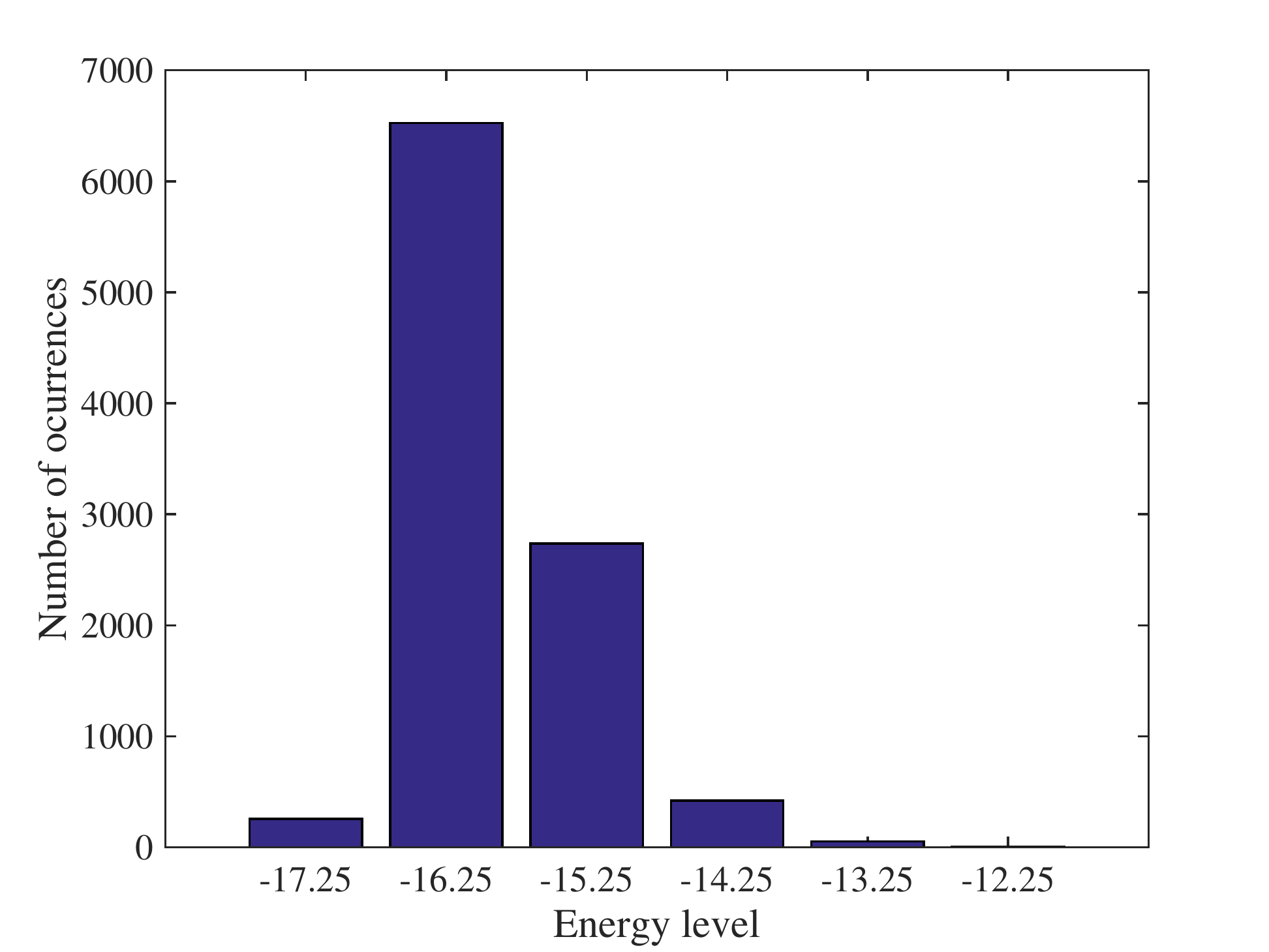}
\caption{Results of the simulation of  the Ising model to find solutions for instance $h_{\Phi}^{\mbox{\tiny qubo}}$. It is shown the frequency of Ising spin configurations of the sampled energy levels using 10000 readouts. Each bar represents the number of times that a configuration with certain energy appears. The sum of all frequencies is equal to the number of requested readouts.}
\label{sec5.fig04}
\end{figure}

\begin{figure}
\centering
\includegraphics[height=6cm]{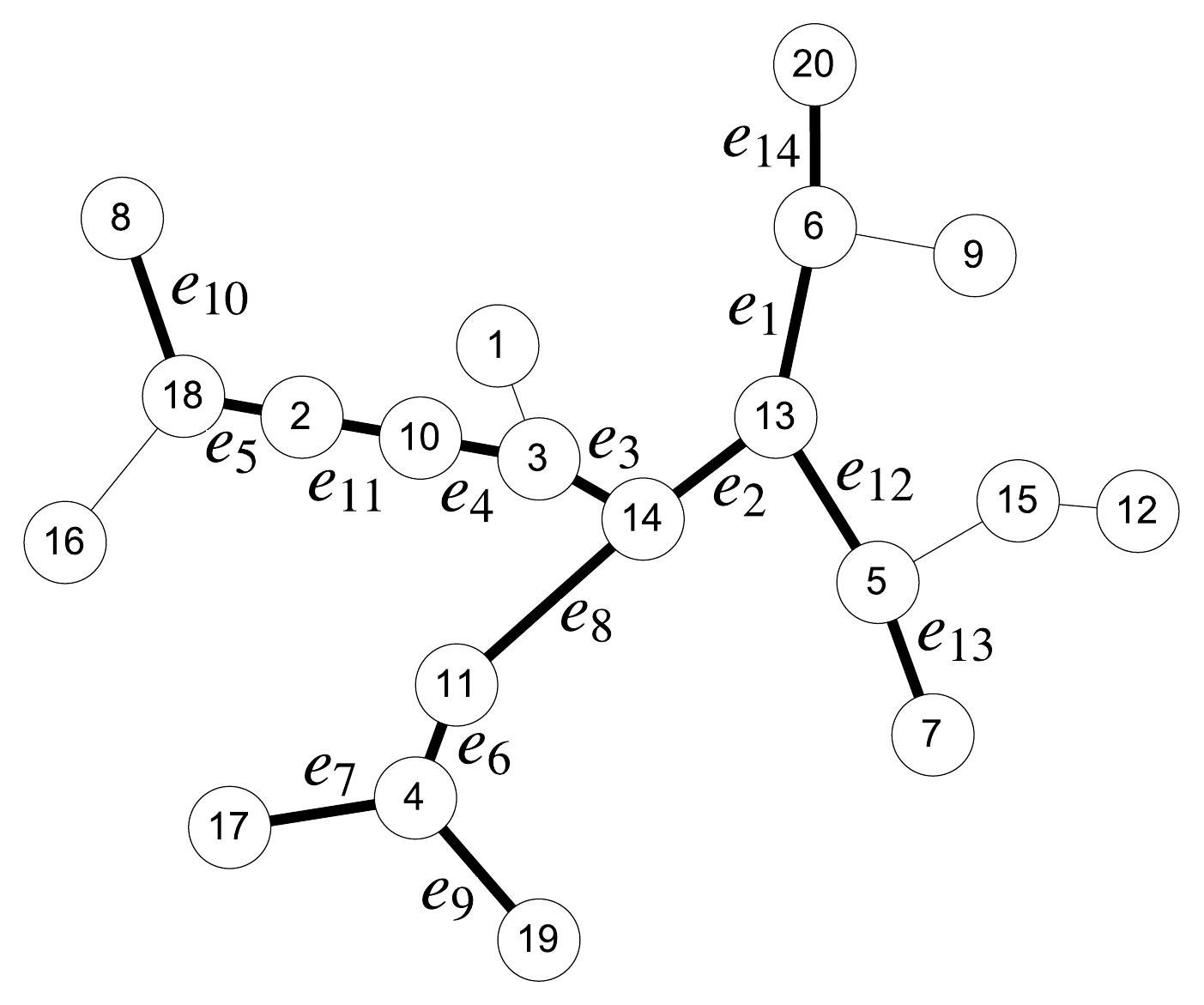}
\caption{A tree instance with 20 vertices for the MMC problem. Bold black lines are the edges that connect the set of pairs in $H=\{(6,10),(2,18),(11,17),(14,19),(8,13),(10,11),(3,5),(13,17),(7,14),(6,20)\}$.}
\label{sec5.fig05}
\end{figure}

\begin{table}[t]
\caption{The first three columns show sampled energy levels, frequencies and Ising spin configurations of the submitted problem depicted in Fig.~(\ref{sec5.fig03}). Only the first 9 spin values of the total 40 Ising variables are shown. The last column shows the number of satisfied clauses in Eq. (\ref{sec5.eq11}), when mapping the spin values -1 to 0 and 1 to 1.}

\label{sec5.tab02}
\centering
\begin{tabular}{llll}
\hline
Energy & Frequency &  Ising configuration & Satisfied  \\ 
             &                   &                        & clauses \\
\hline
-17.25 & 257   & (-1, -1, 1, 1, 1, 1, 1, 1, 1) & 39\\
-16.25 & 6526 & (-1, -1, 1, 1, 1, 1, 1, 1, -1) & 38\\
-15.25 & 2738 & (-1, -1, -1, 1, 1, -1, 1, 1, -1) & 37\\
-14.25 & 422   & (-1, -1, -1, 1, 1, -1, 1, 1, -1) & 37\\
-13.25 & 54     & (-1, -1, -1, 1, 1, -1, 1, 1, 1) & 37\\
-12.25 & 3       & (1, 1, 1, -1, 1, -1, 1, -1, 1) & 36\\

\hline
\end{tabular}
\end{table}

To solve this problem, we use the D-Wave SAPI utility that simulates a quantum computation on a (4,4,4)-Chimera graph with 128 physical qubits. The SAPI utility is a set of software tools designed to mimic the problem-solving behaviour of the hardware solvers. It allows us to use a simulator on a classical computer that solves Ising/QUBO problems by simulated quantum annealing (SQA). It has been proved that the obtained results using SQA have a strong correlation with a D-Wave device~\cite{Boixo2014}. When an Ising problem is submitted to the simulator for a given number of readouts, it returns a list of Ising spin configurations together with their energies, sorted from low to high. The simulator can also return the frequency of each sampled energy level. Fig.~(\ref{sec5.fig04}) shows a histogram of the sampled energies and their frequencies for the Ising problem depicted in Fig.~(\ref{sec5.fig03}), using 10000 readouts. From Table~(\ref{sec5.tab02}), the minimum sampled energy is $E=-17.25 + \text{\em ising\_offset} = -5$ with a frequency of 257 where $\text{\em ising\_offset}=12.25$. The Ising configuration of this energy is $(-1, -1, 1, 1, 1, 1, 1, 1, 1)$ where we omit the other 31 spin values. By mapping the spin value -1 to 0 and 1 to 1, we obtain the assignment $x_1 = 0, x_2 = 0, x_3 = 1, x_4 = 1, x_5 = 1, x_6 = 1, x_7 = 1, x_8 = 1, x_9 = 1$, which satisfies Eq. (\ref{sec5.eq11}) and, therefore, also satisfies the maximum number of clauses of instance  $\Phi$.


The histogram in Fig.~(\ref{sec5.fig04}) also indicates that the probability of measuring the correct answer using 10000 readouts is 257/10000. A D-Wave system also returns approximate solutions with higher energy levels that can be used for other purposes. For instance, the Ising configuration of energy $E=-16.25$ in Table~(\ref{sec5.tab02}) satisfies 38 clauses, one clause below the optimum.

 \subsection{Application to the minimum-multicut problem}

The  {\em Minimum Multicut} (MMC) is an important problem in graph theory that has applications in areas such as telecommunications, routing, VLSI design and circuit partitioning~\cite{Stone1977, Brunetta}. It is defined as follows:

\begin{prob}
Given an undirected graph $G=(V,E)$ and a set of pairs $H=\{(s_1,t_1),\dots,(s_k,t_k)\}\subset V\times V$, find a minimal cardinality subset $E'\subseteq E$ whose removal disconnects each pair in $H$. 
\end{prob}


\begin{figure}
\centering
\includegraphics[height=5cm]{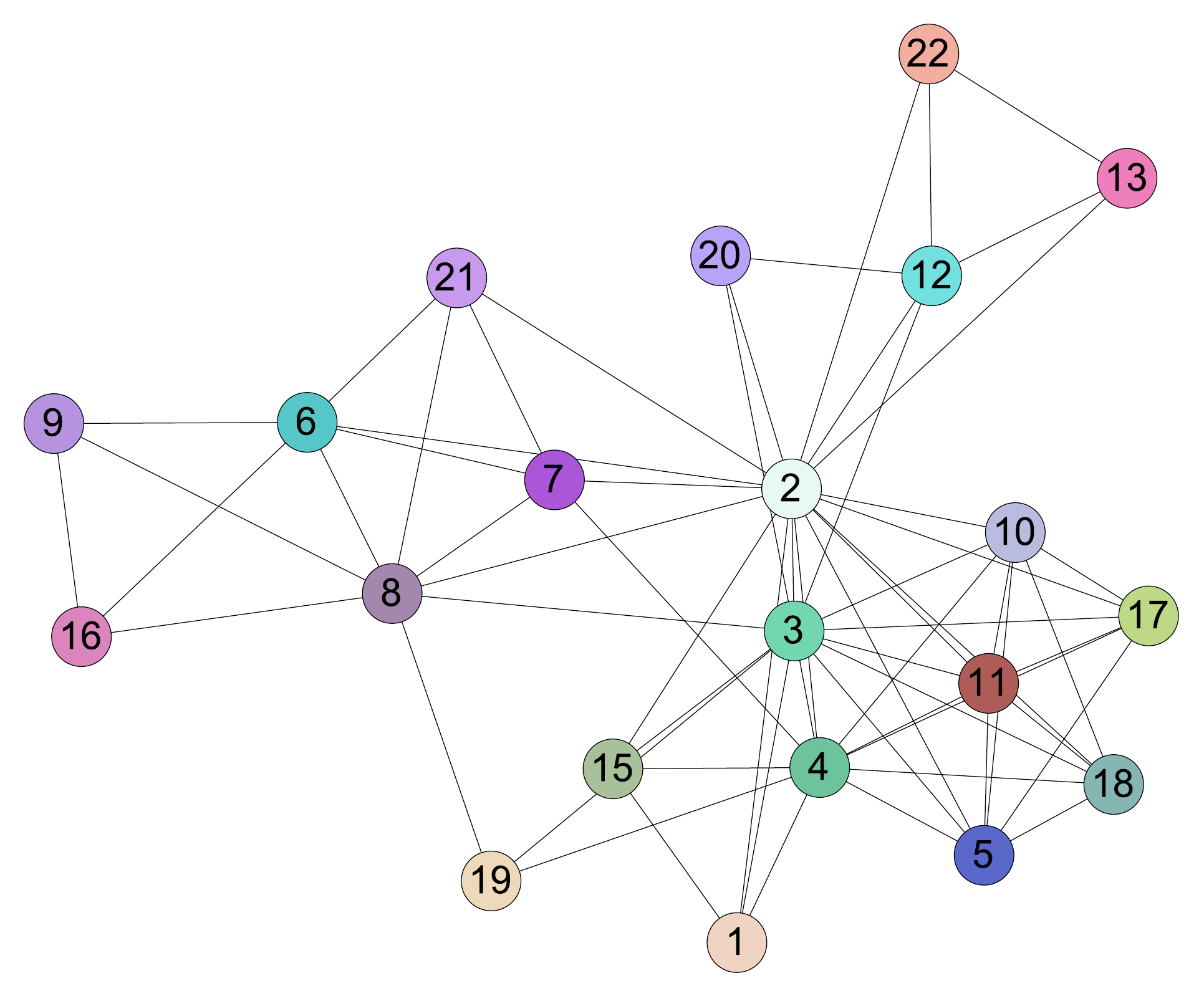}
\includegraphics[height=7cm]{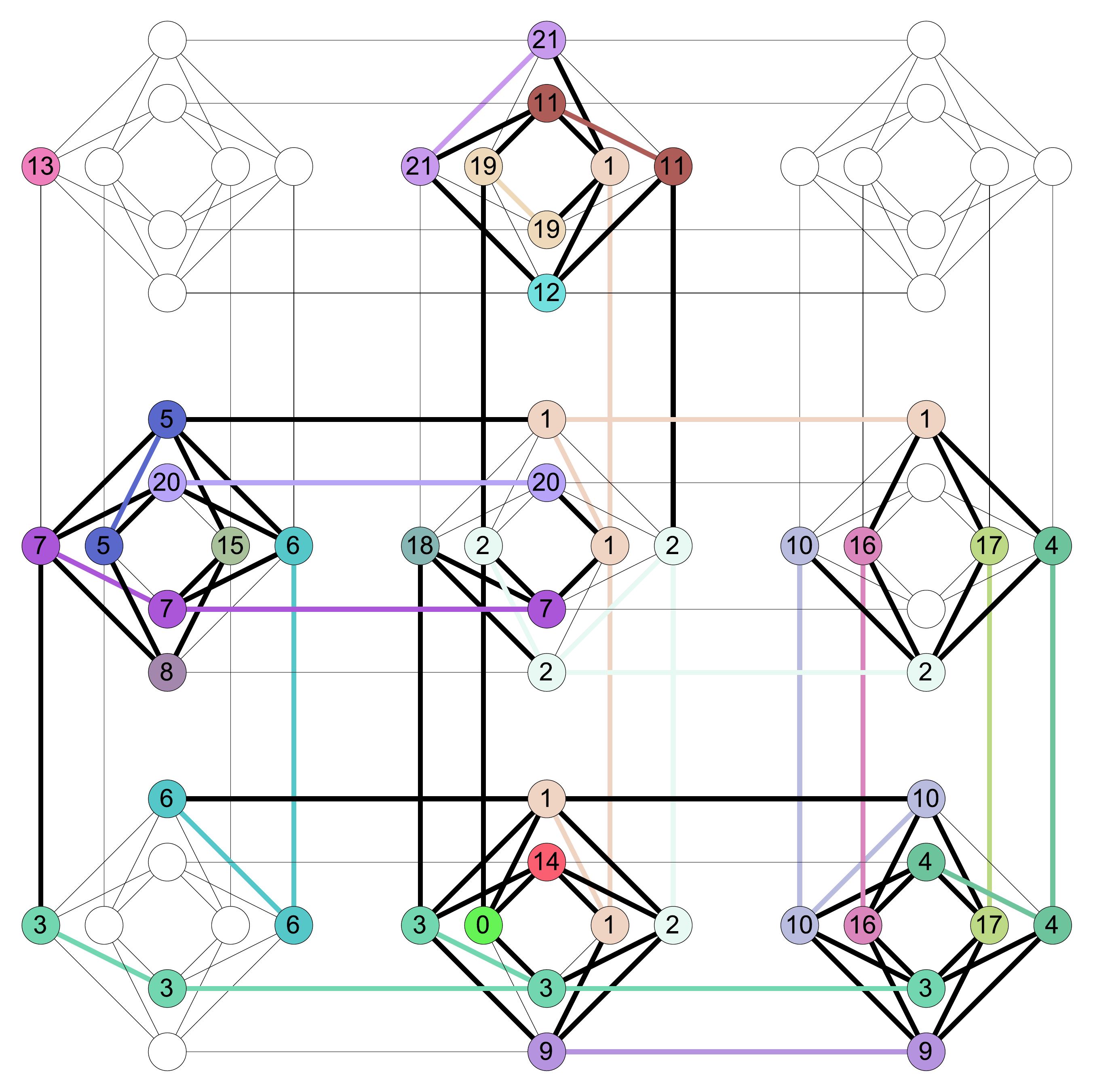} 
\caption{(left) Logical Ising graph for the QUBO function $h_G$ in Eq. (\ref{sec5.eq15}) and (right) a minor embedding of this graph. The logical graph has 23 variables and its minor embedding requires 51 physical qubits. The numbers and colors of the vertices in the logical graph are the same as in the minor embedding. Bold black lines correspond to the mapped edges and bold color lines correspond to chain of qubits.}
\label{sec5.fig06}
\end{figure}

\newpage{}

The MMC problem is $\mathsf{P}$ for $k=2$ and $\mathsf{NP-hard}$ for $k \geq 3$, for general graphs~\cite{DahlhausJPSY94}. In the special case where $G$ is restricted to trees and for arbitrary $k$, the MMC problem remains $\mathsf{NP-hard}$~\cite{Garg1997}.  We will now express the MMC problem in terms of QUBO functions. We consider the case where $G$ is a tree, since it simplifies the formulation of QUBO functions. 


Assuming that $G$ is restricted to a family of trees, then the MMC problem can be formulated as a QUBO function as follows: for each $e\in E$, define a Boolean variable $x_e$ such that $x_e=0$, if $e$ is removed from $G$ and $x_e=1$ otherwise. Let $p_i$ be the unique path between $s_i$ to $t_i$ in $G$ and let $p = \bigcup_{i=1}^k p_i$ be the set of edges in all paths $p_i$. Let

\begin{equation}
h_G = h_{\text{weight}} + h_{\text{penalty}}
\label{sec5.eq14}
\end{equation}
where $h_{\text{weight}} = \sum_{e\in p} (1-x_e)$ and $h_{\text{penalty}} = \lambda_p \sum_{i=1}^k \prod_{e\in p_i} x_e$. The term $h_{\text{weight}}$ coincides with the cardinality of set $C=\{ e\in p | x_e = 0 \}$. Also, $h_{\text{penalty}}=0$ if at least one edge is removed in all paths $p_i$ and $0<h_{\text{penalty}}\leq k\lambda_p$ if at least one path remains connected, where $\lambda_p$ is a positive constant. $\lambda_{\text{path}}$ satisfies that $h_{\text{penalty}}$ adds a penalty value to the function $h_G$ if $C$ is not a multicut in $G$.

Consider an instance of the MMC problem shown in Fig.~(\ref{sec5.fig05}), which consists of a tree with 20 vertices and a set of  disconnected pairs $H=\{(6,10),(2,18),(11,17),(14,19),(8,13),(10,11),(3,5),(13,17),(7,14),(6,20)\}$. The bold lines in Fig.~(\ref{sec5.fig05}) are the edges in the set $p$ of all edges in the paths $p_i$ for every pair $(s_i,t_i) \in H$. For each edge in $p$, we associate a variable $x_j$ for $e_j\in p$ with $j=1,\dots,14$. The function $h_G$ for this instance is

\begin{eqnarray}
h_G &=& 14 - x_1 - x_2 - x_3 - x_4 + 9x_5 - x_6 - x_7 - x_8 - x_9 - x_{10} - x_{11} - \nonumber\\
&& x_{12} -  x_{13} + 9x_{14} + 10x_1 x_2 x_3 x_4 + 10x_6 x_7 + 10x_6 x_8 x_9 + \nonumber\\
&& 10x_2 x_3 x_4 x_5 x_{10} x_{11} + 10x_3 x_4 x_8 + 10x_2 x_3 x_{12} + 10x_2 x_6 x_7 x_8 + \nonumber\\
&& 10x_2 x_{12} x_{13}
\label{sec5.eq15}
\end{eqnarray}
where $\lambda_p=10$ is the number of pairs in $H$. As it can be seen, $\mbox{deg}(h_G)=6$ since the length of the largest path is 6, namely, the path from vertex 8 to 13. There exist seven assignments to variables $x_j $ that produce the minimum cost of five. Using the Ishikawa reduction method, we can obtain a QUBO function $h_G^{\mbox {\tiny qubo}}$ using 9 new variables, given as follows:
\begin{eqnarray}
h_G^{\mbox {\tiny qubo}} &=& 14 - x_{1} - x_{2} - x_{3} - x_{4} + 9x_{5} - x_{6} - x_{7} - x_{8} - x_{9} - x_{10} - x_{11} - x_{12} - \nonumber\\ 
&& x_{13} + 9x_{14} + 20x_{6} x_{7} + 10x_{1} x_{2} + 10x_{1} x_{3} + 10x_{1} x_{4} + 30x_{2} x_{3} + 20x_{2} x_{4} + \nonumber\\
&& 30x_{3} x_{4} + 30x_{15} - 20x_{1} x_{15} - 20x_{2} x_{15} - 20x_{3} x_{15} - 20x_{4} x_{15} + 20x_{6} x_{8} + \nonumber\\
&& 10x_{6} x_{9} + 10x_{8} x_{9} + 10x_{16} - 10x_{6} x_{16} - 10x_{8} x_{16} - 10x_{9} x_{16} + 10x_{2} x_{5} + \nonumber\\
&& 10x_{2} x_{10} + 10x_{2} x_{11} + 10x_{3} x_{5} + 10x_{3} x_{10} + 10x_{3} x_{11} + 10x_{4} x_{5} + 10x_{4} x_{10} + \nonumber\\
&& 10x_{4} x_{11} + 10x_{5} x_{10} + 10x_{5} x_{11} + 10x_{10} x_{11} + 30x_{17} + 70x_{18} - 20x_{2} x_{17} - \nonumber\\
&& 20x_{3} x_{17} - 20x_{4} x_{17} - 20x_{5} x_{17} - 20x_{10} x_{17} - 20x_{11} x_{17} - 20x_{2} x_{18} - \nonumber\\
&& 20x_{3} x_{18} - 20x_{4} x_{18} - 20x_{5} x_{18} - 20x_{10} x_{18} - 20x_{11} x_{18} + 10x_{3} x_{8} + 10x_{4} x_{8} + \nonumber\\
&& 10x_{19} - 10x_{3} x_{19} - 10x_{4} x_{19} - 10x_{8} x_{19} + 20x_{2} x_{12} + 10x_{3} x_{12} + 10x_{20} - \nonumber\\
&& 10x_{2} x_{20} - 10x_{3} x_{20} - 10x_{12} x_{20} + 10x_{2} x_{6} + 10x_{2} x_{7} + 10x_{2} x_{8} + 10x_{7} x_{8} + \nonumber\\
&& 30x_{21} - 20x_{2} x_{21} - 20x_{6} x_{21} - 20x_{7} x_{21} - 20x_{8} x_{21} + 10x_{2} x_{13} + 10x_{12} x_{13} + \nonumber\\
&& 10x_{22} - 10x_{2} x_{22} - 10x_{12} x_{22} - 10x_{13} x_{22}
\label{sec5.eq16}
\end{eqnarray}
and whose logical Ising graph is shown in Fig.~(\ref{sec5.fig06})(left). A minor embedding can be seen in Fig.~(\ref{sec5.fig06})(right) for this logical graph that requires 51 physical qubits.

\begin{figure}[t]
\centering
\includegraphics[height=7cm]{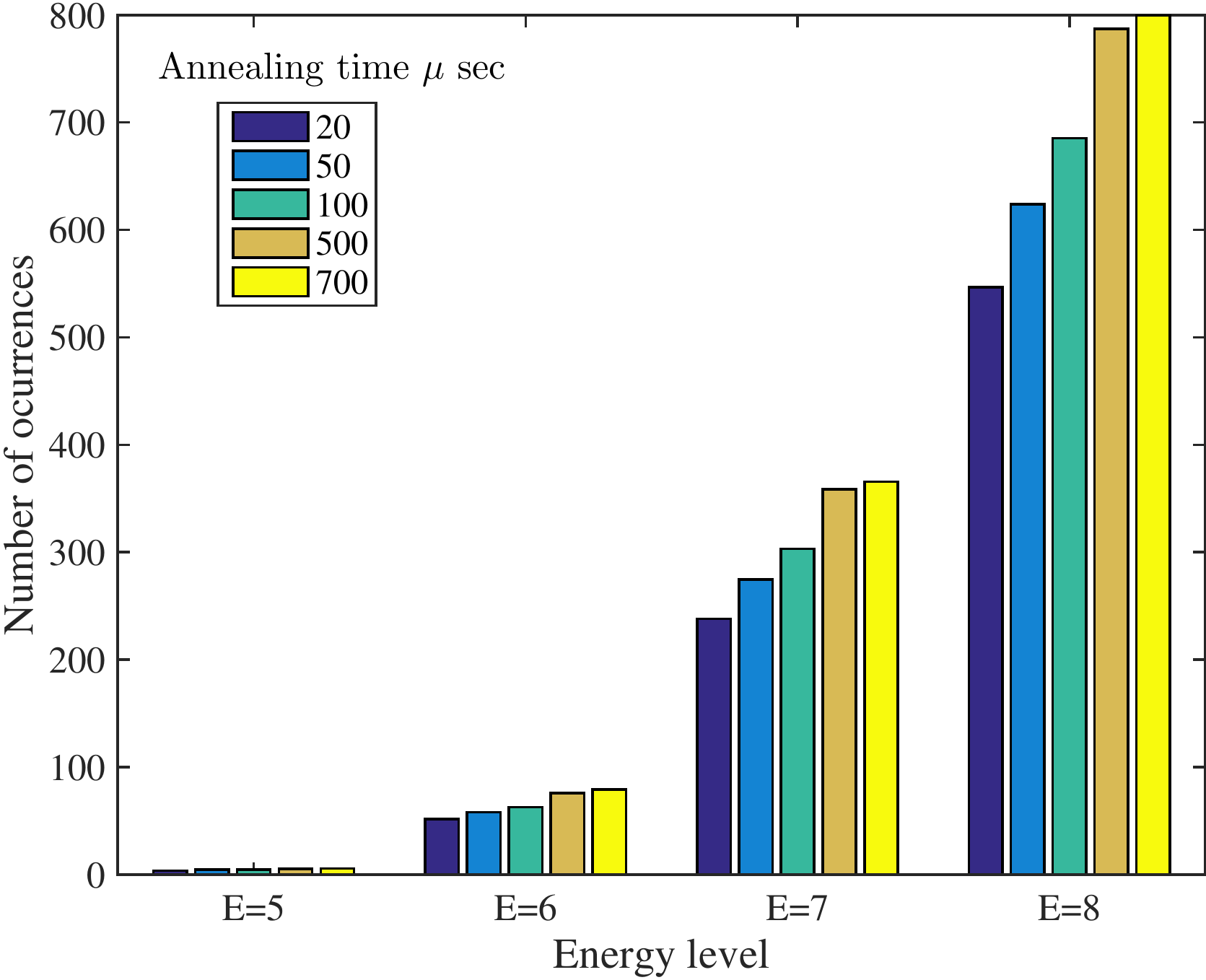}
\caption{Histogram of the first four energy levels for the submitted Ising instance, depicted in Fig.~(\ref{sec5.fig06}), to the D-Wave 2X system. For each energy level, we request a total of 100000 readouts for annealing times $t_a=20, 50, 100, 500, 700 $ microseconds. Each depicted frequency is the average over 100 gauge transformations~\cite{Boixo2014}. A gauge corresponds to a mathematical transformation specified by a vector $a= \{a_1,\dots,a_n\}$, with $a_i = \pm 1$, transforming each Ising variable $s_i\rightarrow a_is_i$. Applying mapping $s_i\rightarrow a_is_i$ to Eq. (\ref{energyfunction}) is equivalent to transform coefficients $h_i\rightarrow a_ih_i$ and $J_{ij}\rightarrow a_ia_jJ_{ij}$. This transformation does not change the landscape energy of the computational problem under analysis, since one can always reverse the sign of the measured spin values $s_i$ given vector ${\bf a}$. However, the experimental results returned by the D-Wave are different, due to calibration errors and precision limitations in the device.}
\label{sec5.fig07}
\end{figure}

As previously said, a D-Wave system receives the annealing time $t_a$ and a number of readouts $N_r$. It returns a histogram of the energies and their Ising spin configurations of the active qubits. Only those Ising spin configurations that correspond to valid cuts have been considered. The Ising problem depicted in Fig.~(\ref{sec5.fig06}) was submitted to the D-Wave 2X system hosted at NASA Ames Research Centre. We requested a total of 100000 readouts in batches of $N_r=1000$ during $R=100$ repetitions. This is because the maximum number of readouts that can be requested is limited according to the chosen annealing time $t_a$. Fig.~(\ref{sec5.fig07}) presents a histogram of the frequencies for the first four lowest sampled energy levels. For each energy level, we show the frequencies for several annealing times. In this experiment, the minimum energy coincides with the number of edges that need to be removed in order to separate every pair in $H$ and whose Ising configuration is (1,-1,-1,1,-1,-1,1,1,1,1,1,1,1,-1). Mapping the spin values from -1 to 0 and 1 to 1, we obtain the assignment (1,0,0,1,0,0,1,1,1,1,1,1,1,0) to the variables $x_1,\dots,x_{14}$. Based on this assignment, the selected edges to form the multicut are $\{e_2,e_3,e_5,e_6,e_{14}\}$ which effectively disconnects every pair in $H$.

Let us estimate the time-to-solution of this instance of the MMC problem on the D-Wave system. Let $p_s$ be the probability of measuring the optimal configuration after a single repetition. Probability $p_s$ can be estimated as the ratio $n_{gs}/N_r$ where $n_{gs}$ is the number of occurrences of the ground state. The number of repetitions $R_P$ to have exactly one success can be formulated as a Bernoulli experiment with probability of success  $P=1-(1-p_s)^{R_P}$ after $R_P$ repetitions. Thus,
\begin{equation}
R_P = \left\lceil \frac{\log(1-P)}{\log(1-p_s)} \right\rceil
\label{sec5.eq17}
\end{equation}
and the time-to-solution $t_{QA}=R_P t_a$. 

Using the data histogram in Fig.~(\ref{sec5.fig07}), we compute $R_P$ and $t_{QA}$ for the average and best frequency over the 100 gauge transformations. In the first case, the average frequency using $t_a=700 \mu s$  is $n_{gs}=5.69$;  the number of repetitions is $R_P=80933$ with a probability $P=0.99$ and the time-to-solution $t_{QA}^{\text{\tiny avg}} = 56.65$ s. In the last case, the best frequency using $t_a=700 \mu s$  is $n_{gs}=91$; the number of repetitions is $R_P=5059$ with a probability $P=0.99$ and the time-to-solution $t_{QA}^{\text{\tiny best}} = 3.5413$ s.
As shown in Fig.~(\ref{sec5.fig04}) and Fig.~(\ref{sec5.fig07}), the frequencies contrast with the simulated data and the experimental data using the D-Wave 2X system. This can be explained by several factors such as the high degree of the function $h_G$ in Eq. (\ref{sec5.eq15}) and the magnitude of the coefficients in the QUBO function $h_G^{\mbox {\tiny qubo}}$, not to mention factors associated with the precision required to represent the problem shown in Eq. (\ref{sec5.eq16}) and the scalability of the number of logical variables and the number of physical qubits, among others.

\section{Conclusions}
\label{conclusions}
This article presents an overview of basic concepts from quantum physics and computer science that are relevant to the study of annealing-based quantum computation and the development of quantum algorithms to attack $\mathsf{NP-complete}$ and $\mathsf{NP-hard}$ problems. 
The components and structure of quantum annealing processors manufactured by D-Wave Systems are presented as well as a discussion about the steps that may be followed in order to build a quantum annealing-based algorithm. Furthermore, we have introduced two detailed examples showing how to solve instances of $\mathsf{NP-hard}$ problems using this novel computing paradigm.

Quantum annealing is a powerful new computational paradigm that explicitly shows the cross-pollination between physics and computer science. While not all problems can be solved in a quantum annealer, it appears that some difficult discrete optimisation problems such as max-SAT and the Minimum Multicut Problem could be computed more efficiently than with traditional classical methods.  


It is only natural to ponder about the potential runtime advantages of quantum annealers over traditional computers to solve practical computational problems (e.g., data mining, machine learning, and route analysis). Unfortunately, it is too early to give a definite answer to the question of computational speed-ups provided by quantum annealers over the best known classical algorithms.   While some promising examples of quantum performance advantages have been found (\cite{trummer,king2015,king2017,ccmClause}),  current generations of D-Wave processors (2000 qubits) are not big enough to solve many $\mathsf{NP-hard}$ applications problems of interesting size (in particular,  keep in mind that embedding into the Chimera is seldom a one-to-one mapping between logical variables and physical qubits).  Additionally, the hardware architecture and programming environment of the D-Wave quantum processors may not be mature enough to fairly compete with the large supercomputers and sophisticated data structures currently available.


Nonetheless, even though the D-Wave quantum annealers cannot be considered at the present time as production run machines, we believe that it is an important platform to conduct research on novel quantum computation techniques and strategies. Furthermore, even though our discussion and experimental results are somewhat specific to D-Wave quantum processors, we believe that they will have a broad impact in the field of annealing- and adiabatic-based quantum computation.

\section*{Acknowledgements}
SEVA, WCS and ML thank USRA for giving us access to the D-Wave quantum computer installed at NASA-Ames Research Centre. Also, SEVA, WCS and ML thank Dr Alejandro Perdomo-Ortiz for useful discussions about programming techniques for D-Wave quantum processors. ML is thankful to Dr Keye Martin for the insightful discussions on adiabatic processes in superconducting qubits at cryogenic temperatures. SEVA warmly thanks his family for their unconditional support.

\section*{Funding}
SEVA gratefully acknowledges the financial support of Tecnologico de Monterrey, Escuela de Ingenieria y Ciencias and CONACyT (SNI number 41594 as well as Fronteras de la Ciencia project No. 1007). 



\begin{thebibliography}{99}


\bibitem{yu2010}
Yu~P, Cardona~M. Fundamentals of semiconductors: physics and materials properties (graduate texts in physics). 4th ed. Berlin:Springer; 2010.

\bibitem{itrs2015}
ITRS.  International Technology Roadmap for semiconductors 2.0 executive report. 2015.


\bibitem{wong2002}
Wong~H-SP. Beyond the conventional transistor. IBM J. Res. \& Dev. 2002 Mar;46(2/3):133--168.

\bibitem{wong2010} 
Wong~B, Mittal~A, Cao~Y, Starr~GW. Nano-CMOS Circuit and Physical Design. 1st ed.: Hoboken (NJ):Wiley \& Sons; 2005.

\bibitem{avci2014}
Avci~UE, Morris~DH, Young~IA. Tunnel Field-Effect Transistors: Prospects and Challenges. IEEE J. Electron Devices Soc 2015;3(3):88--95.

\bibitem{kirkpatrick83}
Kirkpatrick~S, Gelatt Jr~CD, Vecchi~MP. Optimization by Simulated Annealing. Science 1983;220(4598): 671--680.

\bibitem{baez12}
Baez~J, Stay~M. Algorithmic thermodynamics. Mathematical Structures in Computer Science 2012; 22(5):771--787.


\bibitem{compeau12} 
Compeau~P, Pevzner~P. Bioinformatics Algorithms: An Active Learning Approach, 2nd Ed. Vols. 1 \& 2.  Active Learning Publishers (2015)

\bibitem{cubitt15}
Cubitt~TS, Perez-Garcia~D, Wolf~MM. Undecidability of the spectral gap. Nature 2015;528(7581):207--211.

\bibitem{wittek14}
Wittek~P. Quantum Machine Learning. Academic Press (2014)

\bibitem{schulda15}
Schulda~M, Sinayskiy~I, Petruccione~F. An introduction to quantum machine learning. Contemp. Phys. 2015;56(2):172--185.

\bibitem{biamonte17}
Biamonte~J, Wittek~P, Pancotti~N, Rebentrost~P, Wiebe~N, Lloyd~S. Quantum machine learning. Nature. 2017;549:195--202.


\bibitem{yan15}
Yan~F, Iliyasu~AM, Venegas-Andraca~SE. A survey of quantum image representations. Quantum Information Processing 2016;15(1):1--35. 

\bibitem{aburaed17} Abura'ed~N, Khan~FS, Bhaskar~H. Advances in the Quantum Theoretical Approach to Image Processing Applications. ACM Comput. Surv. 2017; 49(4):1--49.

\bibitem{idquantique} ID Quantique. http://www.idquantique.com/

\bibitem{dwave} D-Wave systems. http://www.dwavesys.com/

\bibitem{ibm} IBM Quantum Experience. http://research.ibm.com/ibm-q/

\bibitem{microsoft} Microsoft Station Q. https://stationq.microsoft.com/

\bibitem{googlequantum} Google Quantum Artificial Intelligence Laboratory. https://plus.google.com/+QuantumAILab/

\bibitem{1qbit} 1Qbit. http://1qbit.com/

\bibitem{rigetti} Rigetti. http://rigetti.com/

\bibitem{qmanifesto} Quantum Manifesto, an EU  call to invest \euro 1 billion on  quantum technologies. 

\bibitem{ukquantumhubs} UK National Quantum Technologies Programme. http://uknqt.epsrc.ac.uk/

\bibitem{qis_executive_usa} USA National Science and Technology Council. Advancing Quantum Information Science: National Challenges and Opportunities (July 2016). 


\bibitem{mittechreviewquantum2017} MIT Tech Review 10 Breakthrough Technologies 2017 - Practical Quantum Computers. 

\bibitem{winiarczyk13} Winiarczyk~R, Gawron~P, Miszczak~JA, Pawela~\L{}, Pucha\l{}a~Z. Analysis of patent activity in the field of quantum information processing. Int. J. Quantum Inform. 2013; 11: 1350007.

\bibitem{ukquantumtech14} UK Intellectual Property Office Informatics Team. Eight great technologies. Quantum Technologies, a patent overview. UK Intellectual Property Office (2014). 

\bibitem{usclockheedmartin} 
 USC-Lockheed Martin Quantum Computation Center. https://www.isi.edu/research\_groups/quantum\_computing/home
\bibitem{hpquantumlabs} 
HP Labs Quantum Information Processing Group. http://www.hpl.hp.com/research/qip/

\bibitem{mertens02}
Mertens~S. Computational Complexity for Physicists. Computing in Science \& Engineering 2002;4(3):31--47. 

\bibitem{press92}
Press~WH, Flannery~BP, Teukolsky~SA, Vetterling~WT. Numerical Recipes in C: The Art of Scientific Computing, 2nd Edition. Cambridge: Cambridge University Press; 1992.

\bibitem{neosguide}
NEOS Guide. https://neos-guide.org/content/combinatorial-optimization

\bibitem{savage98} Savage~JE. Models of Computation. Addison-Wesley Publishers; 1998.

\bibitem{garey79} Garey~MR, Johnson~DS. Computers and intractability. A guide to the Theory of NP-Completeness. W.H. Freeman and Company; 1979.

\bibitem{du16} Du~K-L, Swamy~M.N.S. Search and Optimization by Metaheuristics. Birkh\"auser; 2016.


\bibitem{guenin15} Guenin~B, K\"onemann~J, Tun\c{c}el~L. A Gentle Introduction to Optimization. Cambridge University Press;2015.



\bibitem{sipser06} Sipser~M. Introduction to the theory of computation. Thomson Learning Inc.; 2006.

\bibitem{papadimitriou95} Papadimitriou~CH. Computational complexity. Addison-Wesley; 1995.

\bibitem{moore11} Moore~C, Mertens~S. The nature of computation. Oxford University Press; 2011.

\bibitem{goldreich10} Goldreich~ O. P, NP, and NP-completeness. Cambridge University Press; 2011.

\bibitem{marquessilva08}
Marques-Silva~J. Practical Applications of Boolean Satisfiability. IEEE 9th International Workshop on Discrete Events Systems 2008;74--80.


\bibitem{complexityzoo16} The Complexity Zoo. https://complexityzoo.uwaterloo.ca



\bibitem{deutsch85} 
Deutsch~D. Quantum theory, the Church-Turing principle and the universal quantum computer. Proc. R. Soc. Lond. A 1985;400(1818):97--117.

\bibitem{strassen69}
Strassen~V.  Gaussian Elimination is not Optimal, Numer. Math. 1969;13:354--356.

\bibitem{hedtke11}
Hedtke~I. Strassen's Matrix Multiplication Algorithm for Matrices of Arbitrary Order. Bulletin of Mathematical Analysis and Applications 2011; 3(2):269--277.

\bibitem{playground} http://www.quantumplayground.net/

\bibitem{liquid} http://stationq.github.io/Liquid/

\bibitem{qmdirac} http://qmdirac.weebly.com/

\bibitem{turing36} 
Turing~A.M. On Computable Numbers, with an application to the Entscheidung problem. Proceedings of the London Mathematical Society 1936-37;42:230--265. 

\bibitem{knuth74}
Knuth~DE.  A terminological proposal, ACM SIGACT News. January 1974;6(1):12--18.

\bibitem{cook71}
 Cook~S. The complexity of theorem proving procedures. Proceedings of the Third Annual ACM Symposium on Theory of Computing;1971:151--158.

\bibitem{levin73}
  Levin~L. Universal search problems. Problems of Information Transmission (in Russian) 1973;9(3):115--116.

\bibitem{henderson03}
Henderson~D, Jacobson~SH and Alan W. Johnson. Handbook of Metaheuristics. Boston:Springer; 2003. Chapter 10, The Theory and Practice of Simulated Annealing: 287--319.

\bibitem{kadowaki98}
Kadowaki~T, Nishimori~H. Quantum annealing in the transverse Ising model.  Phys. Rev. E 1998;58(5): 5355--5363.

\bibitem{battaglia06}
Battaglia~DA, Stella~L. Optimization through quantum annealing: theory and some applications.  Contemp. Phys. 2006;47(4):195--208.

\bibitem{bapst11}
Bapst~V, Semerjian~G. Thermal, quantum and simulated quantum annealing: analytical comparisons for simple models.  The European Physical Journal Special Topics 2011;473:1--8 (012011).

\bibitem{martonak02} Marto\v n\'ak~R, Santoro~GE, Tosatti~E. Phys. Rev. E 2004;58: 057701.

\bibitem{crosson2017} Crosson~E, Harrow~QW. Simulated quantum annealing can be exponentially faster than classical simulated annealing. Proceedings of the 57th IEEE Annual Symposium on Foundations of Computer Science;2017:714--723.

\bibitem{galvao03} 
Galv\~ao~EF, Hardy~L. Substituting a Qubit for an Arbitrarily Large Number of Classical Bits. Phys. Rev. Lett. 2003;90:087902.


\bibitem{sornborger12} 
Sornborger~AT. Substituting a Qubit for an Arbitrarily Large Number of Classical Bits. Sci. Rep. 2012;2:597.



\bibitem{bian-csp} Bian~Z, Chudak~F, Israel~R, Lackey~B, Macready~WG, Roy~A. Mapping constrained optimization problems to quantum annealing with applications to fault diagnosis.  Frontiers in ICT;28 July 2016:https://doi.org/10.3389/fict.2016.00014. 

 
\bibitem{berkley} Berkley~AJ, Johnson~MW, Bunyk~P, Harris~R, Johansson~J, Lanting~T, Ladizinsky~E, Tolkacheva~E, Amin~MHS, Rose~G. A scalable readout system for a superconducting adiabatic quantum optimization system. Superconductivity Sci. Tech. 2010; 23(10):105014. 

\bibitem{bunyk} Bunyk~PI, Hoskinson~E, Johnson~MW, Tolkacheva~I. Architectural considerations in the design of a superconducting quantum annealing processor. IEEE Transactions on Applied Superconductivity 2014;24(4):1700110.

\bibitem{choi1} Choi~V. Minor-embedding in adiabatic quantum computation: I. The parameter setting problem. Quantum Information Processing 2008;7(5):  193--209. 

\bibitem{choi2} Choi~V.  Minor-embedding in adiabatic quantum computation: II. Minor-universal graph design. Quantum Information Processing 2011; 10(3): 343--353. 

\bibitem{denchev} Denchev~VS, Boixo~S, Isakov~SV, Ding~N, Babbush~R, Smelyanskiy~V, Martinis~J, Neven~H. What is the computational value of finite range tunneling?.  Phys. Rev. X 2015; 6(3): 031015.  

\bibitem{dickson} Dickson~NG, et~al. Thermally assisted quantum annealing of a 16-qubit problem. Nature Communications 2013;4:1903.  

\bibitem{harris} Harris~R, et~al. Experimental demonstration of a robust and scalable flux qubit. Phys. Rev. B 2010;81:134510.



\bibitem{lucas-formulations}  Lucas~A. Ising formulations of many NP problems. Frontiers in Physics 2014;  2(5): https://doi.org/10.3389/fphy.2014.00005.

\bibitem{mcgeoch-qa} McGeoch~CC.  Adiabatic Quantum Computation and Quantum Annealing:  Theory and Practice. Synthesis Lectures in Quantum Computing,  Morgan and Claypool; 2014.  


\bibitem{mcgeoch-wang} McGeoch~CC, Wang~C. Experimental evaluation of an adiabatic quantum system for combinatorial optimization. Proc. 2013 Conf. on Computing Frontiers,  ACM; 2013.  




\bibitem{trummer} Trummer~I, Koch~C. Multiple query optimization on the D-Wave 2X adiabatic quantum computer.  Proceedings of the 42nd Intl. Conf. on Very Large Ddata Bases (VLDB); 2016. 


\bibitem{mwesigwa} Ushijima-Mwesigwa~H, Negre~CFA , Mniszewski~SM. Graph partitioning using quantum annealing on the D-Wave system. arXiv:1705.03082v1;  4 May 2017.   


\bibitem{venturelli-jobshop}  Venturelli~D, Marchand~DJJ, Roho~G. Quantum annealing implementation of Job-Shop Scheduling. Proceedings of the 26th International Conference on Automated Planning and Scheduling; 2016.   



\bibitem{Kaminsky2004a}
Kaminsky~WM, Lloyd~S. Scalable architecture for adiabatic quantum computing of NP-hard problems. Quantum Computing and Quantum Bits in Mesoscopic Systems 2004; Springer:229-236.

\bibitem{Kaminsky2004b}
Kaminsky~WM, Lloyd~S, Orlando~TP. Scalable superconducting architecture for adiabatic quantum
computation. arXiv.org:quant-ph/0403090, 2004.

\bibitem{Boros2002}
Boros~E, Hammer~PL. Pseudo-boolean Optimization. Discrete Appl. Math. 2002;123(1-3):155--225.

\bibitem{Freedman05}
Freedman~D, Drineas~P. Energy minimization via graph cuts: settling what is possible. CVPR 2005: 939--946.

\bibitem{Ishikawa11}
Ishikawa~H. Transformation of General Binary MRF Minimization to the First-Order Case. IEEE Trans. Pattern Anal. Mach. Intell. 2011;33(6):1234--1249.

\bibitem{Stone1977}
Stone~HS. Multiprocessor Scheduling with the Aid of Network Flow Algorithms. IEEE Transactions on Software Engineering 1977;3(1):85--93.

\bibitem{Brunetta}
Brunetta~L, Conforti~M, Fischetti~M. A polyhedral approach to an integer multicommodity flow problem. Discrete Applied Mathematics 2000;101(1-3):13- 36.

\bibitem{DahlhausJPSY94}
Dahlhaus~E, Johnson~DS, Papadimitriou~CH, Seymour~PD,Yannakakis~ M. The Complexity of Multiterminal Cuts. SIAM J. Comput. 1994;23(4):864--894.

\bibitem{Garg1997}
Garg~N, Vazirani~VV, Yannakakis~ M. Primal-dual approximation algorithms for integral flow and multicut in trees. Algorithmica 1997;18(1):3--20.


\bibitem{AusielloBook}
Ausiello~G,  Protasi~M, Marchetti-Spaccamela~A, Gambosi~G, Crescenzi~P,  Kann~V. Complexity and Approximation: Combinatorial Optimization Problems and Their Approximability Properties. Springer-Verlag New York, Inc.;1999.


\bibitem{Boros2014}
Boros~E, Gruber~ A. On Quadratization of Pseudo-Boolean Functions. International Symposium on Artificial Intelligence and Mathematics 2014.


\bibitem{Anthony20161}
Anthony~M, Boros~E, Crama~Y, Gruber~A. Quadratization of symmetric pseudo-Boolean functions. Discrete Applied Mathematics 2016;2031--12.

\bibitem{Boixo2014}
 Boixo~S, R{\o}nnow~TF, Isakov~SV, Wang~Z, Wecker~D, Lidar~DA, Martinis~JM, Troyer~ M. Evidence for quantum annealing with more than one hundred qubits. Nature Physics 2014;10:218--224.


\bibitem{arora09}
Arora~S, Barak~B. Computational Complexity: a modern approach. Cambridge University Press; 2009.



\bibitem{king2015}  King~J. et~al., Benchmarking a quantum annealing processor with the time-to-target metric.   arXiv:1508.05087v1. 2015.  

\bibitem{king2017} King~J. et~al., Quantum annealing amid local ruggedness and global frustration.  arXiv:1701.04579v2. 2017.  

\bibitem{ccmClause} C. C. McGeoch, {\em Optimization with Clause Problems}, D-Wave Technical Report  Series No. 14-1001A-A. 2017 



\end{thebibliography}
\end{document}